\documentstyle[12pt,a4,epsf]{article}

\def\tr{\mathop{\rm tr}\nolimits}

\makeatletter
\@addtoreset{equation}{section}
\makeatother

\newcommand{\VEV}[1]{\left\langle #1 \right\rangle}

\newcommand{\nn}{\nonumber}

\newcommand{\bequ}{\begin{equation}}
\newcommand{\eequ}{\end{equation}}
\newcommand{\beqn}{\begin{eqnarray}}
\newcommand{\eeqn}{\end{eqnarray}}
\newcommand{\Ls}{\left(}
\newcommand{\Rs}{\right)}
\newcommand{\Lm}{\left\{}
\newcommand{\Rm}{\right\}}
\newcommand{\Ll}{\left[}
\newcommand{\Rl}{\right]}

\begin{document}
\begin{titlepage}

\begin{flushright}
hep-ph/0202050\\
KUNS-1754\\
\today
\end{flushright}

\vspace{4ex}

\begin{center}
{\large \bf
$E_6$ Unification, Doublet-Triplet Splitting \\
and  Anomalous $U(1)_A$ Symmetry
}

\vspace{6ex}

\renewcommand{\thefootnote}{\alph{footnote}}
Nobuhiro Maekawa\footnote
{E-mail: maekawa@gauge.scphys.kyoto-u.ac.jp
}
and 
Toshifumi Yamashita\footnote{
E-mail: yamasita@gauge.scphys.kyoto-u.ac.jp
}

\vspace{4ex}
{\it Department of Physics, Kyoto University, Kyoto 606-8502, Japan}\\
\end{center}

\renewcommand{\thefootnote}{\arabic{footnote}}
\setcounter{footnote}{0}
\vspace{6ex}

\begin{abstract}
We propose a natural Higgs sector in $E_6$ grand unified theory
(GUT) with anomalous $U(1)_A$ gauge symmetry. In this scenario, 
the doublet-triplet splitting can be realized, 
while proton decay via dimension 5 operators is suppressed. 
Gauge coupling unification is also realized without fine-tuning.
The GUT scale obtained in this scenario is generally lower than 
the usual GUT scale, $2\times 10^{16}$ GeV, and therefore it should 
be possible to observe proton decay via dimension 6 operators
in near future experiments.
The lifetime of a nucleon in this model is roughly estimated as 
$\tau_p(p\rightarrow e^+\pi^0)\sim 3\times 10^{33}$ years. 
It is shown that the Higgs sector is compatible with the matter sector
proposed by one of the present authors, 
which reproduces realistic quark and lepton mass matrices, including
a bi-large neutrino mixing angle. Combining the Higgs sector
and the matter sector, we can obtain a completely consistent $E_6$ GUT.
The input parameters for this model are only eight integer 
anomalous $U(1)_A$ charges (+3 for singlet Higgs) for 
the Higgs sector and three (half) integer charges for the matter sector.

\end{abstract}

\end{titlepage}

\section{Introduction}
Recently in a series of papers
\cite{maekawa,maekawa2,maekawa3} a scenario to construct a realistic 
 GUT  has been 
proposed within the $SO(10)$ group. In this scenario, anomalous $U(1)_A$
gauge symmetry \cite{U(1)}, whose anomaly is cancelled by the 
Green-Schwarz mechanism \cite{GS}, plays a critical role, and 
it has many interestingfeatures:
1) the interaction is generic, in the sense that all the interactions 
that are allowed by the symmetry are introduced. Therefore, once we 
fix the field content with its quantum numbers~(integer), all the 
interactions are determined, except the coefficients of order 1;
2) it naturally solves the so-called 
doublet-triplet (DT) splitting problem \cite{DTsplitting}, 
using the Dimopoulos-Wilczek~(DW) mechanism
\cite{DW,BarrRaby,Chako,complicate};  
3) it reproduces realistic structure of quark and lepton 
mass matrices, including neutrino bi-large mixing \cite{SK},
 using the Froggatt-Nielsen (FN) mechanism\cite{FN};  
4) the  anomalous $U(1)_A$ explains the hierarchical structure of 
the symmetry-breaking scales  and the masses of 
heavy particles; 
5) all the fields, except those of the minimal 
supersymmetric standard model (MSSM), can become heavy;
6) the gauge couplings are unified just below the usual GUT scale 
$\Lambda_G\sim 2\times 10^{16}$ GeV;
7) in spite of the lower GUT scale, proton decay via dimension 6 
operators, $p\rightarrow e^+\pi^0$, is within experimental 
limits, leading us to expect that proton decay will soon be observed.
8) the cutoff scale is lower than the Planck scale;
9) the $\mu$ problem is also solved.

An extension of the above mentioned $SO(10)$ model 
to the $E_6$ gauge group has been carried out
 in the analysis of fermion masses~\cite{BM}, 
and it has been found that $E_6$ is more economical in the sense 
that we have only to introduce minimal matter fields 
three ${\bf 27}$s  for 3-family fermions in contrast to the 
matter content of three ${\bf 16}$ plus one ${\bf 10}$ in 
the $SO(10)$ case. 
Moreover, the charge assignment for realizing bi-large
neutrino mixing automatically satisfies the condition for weakening 
the flavor changing neutral current (FCNC). 
Specifically, the right-handed down quark and the left-handed
lepton of the first and second generations belong to a single 
multiplet ${\bf 27}$ as a result of the twisting family structure 
\cite{Bando}.

In the Higgs sector of $E_6$ unification, however, 
the situation is not so simple. The Higgs fields 
$\bar \Phi({\bf \overline{27}})$ and $\Phi({\bf 27})$ are needed to
break $E_6$ into $SO(10)$, in addition to the Higgs fields 
$A({\bf 78})$, whose VEV breaks $SO(10)$ into 
$SU(3)_C\times SU(2)_L\times SU(2)_R\times U(1)_{B-L}$, 
and $\bar C({\bf \overline{27}})$ and  $C({\bf 27})$, whose VEVs break
$SU(2)_R\times U(1)_{B-L}$ into $U(1)_Y$.
The other VEVs, 
$\VEV{\bar C}$, $\VEV{C}$, $\VEV{\bar\Phi}$ and $\VEV{\Phi}$, 
generally destabilize the DW-type VEV, 
$\VEV{{\bf 1}_A}=\VEV{{\bf 16}_A}=\VEV{{\bf \overline{16}}_A}=0$, 
$\VEV{{\bf 45}_A}=\tau_2\times {\rm diag}
(v,v,v,0,0)$, which is required to realize 
DT splitting. Therefore we may have to remove 
the interaction between $A$ and $\bar\Phi\Phi$ as well as that 
between $A$ and $\bar CC$. However, if we forbid 
interactions between these Higgs fields in the superpotential,
then pseudo-Nambu-Goldstone (PNG) fields appear. 
Since the non-vanishing VEVs 
$|\VEV{\Phi({\bf 1,1})}|=|\VEV{\bar \Phi({\bf 1,1})}|$ break
$E_6$ into $SO(10)$, the Nambu-Goldstone (NG) modes 
${\bf 16}$, ${\bf \overline{16}}$ and ${\bf 1}$ of $SO(10)$ appear.
Also, the VEVs 
$|\VEV{C({\bf 16,1})}|=|\VEV{\bar C({\bf \overline{16},1})}|$ break 
$E_6$ into another $SO(10)'$, so 
${\bf 16}$, ${\bf \overline{16}}$ and ${\bf 1}$ of $SO(10)'$ again 
appear as NG modes. 
Here, the representations of $SO(10)$ and $SU(5)$ are explicitly 
denoted as the first and second numbers in the parentheses, 
respectively.
In the language of usual $SU(5)$,
these NG modes are represented as $({\bf 10}+{\bf \bar 5}+{\bf 1})$, 
$({\bf \overline{10}}+{\bf 5}+{\bf 1})$ and ${\bf 1}$.
The VEV of the adjoint field 
$\VEV{{\bf 45}_A}=\tau_2\times {\rm diag}(v,v,v,0,0)$
breaks $E_6$ into $SU(3)_C\times SU(2)_L\times SU(2)_R\times U(1)_{B-L}
\times U(1)_{V'}$, and 
the NG modes resulting from  this breaking are ${\bf 16}$ and 
${\bf \overline{16}}$
 of $SO(10)$ and $({\bf 3,2})_{\frac{1}{6}}+
({\bf \bar 3,1})_{-\frac{2}{3}}+({\bf 3,2})_{-\frac{5}{6}}+h.c.$
of the standard gauge group 
$SU(3)_C\times SU(2)_L\times U(1)_Y$. 
Since some of these modes are absorbed by the Higgs mechanism,
the remaining PNG modes become
${\bf 10+\overline{10}}+2\times {\bf 5}+2\times{\bf \bar 5
}+4\times {\bf 1}$ of $SU(5)$
and $({\bf 3,2})_{\frac{1}{6}}+({\bf \bar 3,1})_{-\frac{2}{3}}+h.c.$
of the standard gauge group. If all these PNG modes have only tiny 
masses around the weak scale, then not only is coupling unification 
destroyed but also the gauge couplings diverge below the GUT scale.
Therefore, we have to give these PNG fields superheavy masses.
However, in order to do so, we have to introduce
some interactions between Higgs fields, and this requirement is 
in opposition to that needed 
to stabilize the DW-type of VEV. This conflict is similar to that
existing in the $SO(10)$ case. 

This paper aims at obtaining a unified description of the Higgs 
sector in $E_6$ model, in which 
the above stated problem and the DT splitting problem are solved.
It may seem that $SO(10)$ unified models are promising for this purpose.
However,  if we proceed to $E_6$, there are more advantages
in addition to the natural FCNC suppression. 
In particular, we have the following:
\begin{enumerate}
\item  The FN field 
naturally emarges as the composite operator $\VEV{\bar \Phi\Phi}$,
where $\Phi$ and $\bar \Phi$  are  needed to break 
$E_6$ down to $SO(10)$.
\item The usual doublet Higgs field $H$ is already included 
in the field 
$\Phi$.
\item In the Higgs sector, the condition 
for the unification of gauge 
coupling constants automatically provides ``R parity" 
in terms of anomalous $U(1)_A$ naturally, and therefore we do not 
have to introduce additional R parity.
\end{enumerate}
Moreover, we can construct a completely consistent and realistic
$E_6$ GUT scenario by combining this Higgs sector and 
the matter sector.

After explaining how the vacuum in the Higgs sector is determined
by anomalous $U(1)_A$ charges (\S2) and 
giving a quick review of the $SO(10)$ 
model~(\S3), we explain how the above desirable features in the Higgs 
sector are naturally obtained in the $E_6$ unification (\S4) and
a completely consistent $E_6$ GUT scenario~(\S5).

\section{Vacuum determination}
Here we explore some general structures of VEVs that are 
determined from the superpotential of the Higgs sector. 
The Higgs sector is the most poorly part, and usually the VEVs of 
Higgs fields are introduced as input parameters, because
general forms of the potential are usually too arbitrary. 
However, anomalous $U(1)_A$ provides us with very strong constraints 
on the superpotential  $W$  and thus dictates the scales of the system 
in a quite definite way.   
In our analysis, supersymmetry (SUSY) is essential, because the 
observation of the following vacuum structure is due to   
the analytic property of SUSY theory. 

We study the simplest case, 
in which  all the fields
are gauge singlet fields 
$Z_i^\pm$ ($i=1,2,\cdots n_\pm$)
 with charges $z_i^\pm$ ($z_i^+>0$ and $z_i^-<0$).
Through out this paper  we use units in which the cutoff 
is $\Lambda=1$, and we denote all the
superfields with uppercase letters and their anomalous $U(1)_A$ charges
with the corresponding lowercase letters.

We first show that none of the fields with positive anomalous
$U(1)_A$ charge acquire non-zero VEV if the FN
mechanism
\cite{FN}
acts effectively in the vacuum.
From the $F$-flatness conditions of the superpotential,
we get $n=n_++n_-$ equations  plus one $D$-flatness condition,
\begin{equation}
 F_{Z_i}\equiv\frac{\partial  W}{\partial Z_i}=0, \qquad D_A=g_A
      \left(\sum_i z_i |Z_i|^2 +\xi^2 \right)=0,
\label{eq:fflat}
\end{equation}
where 
$\xi^2$
is the coefficient of the Fayet-Iliopoulos (FI) $D$-term.
\footnote{
In weakly coupled Heterotic string theory, this FI $D$-term can be
induced by a stringy loop correction, according to which 
$\xi^2=\frac{g_s^2\tr Q_A}{192\pi^2}$.}
The above equations may seem to be over determined. However,
 the $F$-flatness
conditions are not independent, because of the gauge invariance 
expressed by 
\begin{equation}
\frac{\partial  W}{\partial Z_i}z_iZ_i=0.
\label{constraint}
\end{equation}
Therefore, generically a SUSY vacuum with $\VEV{Z_i}\sim \Lambda$ 
exists
(Vacuum A),
because the coefficients of the terms of $F$-flatness
conditions are generically
of order 1. 
However, if $n_+\leq n_-$, we can choose another vacuum (Vacuum B)
with $\VEV{Z_i^+}=0$, which automatically satisfy the $F$-flatness
conditions $F_{Z_i^-}=0$, since they contain
at least one field with positive charge. Then, the $\VEV{Z_i^-}$
are determined by the $F$-flatness conditions
$F_{Z_i^+}=0$, with the constraint
(\ref{constraint}), and the $D$-flatness condition $D_A=0$.
Note that if $\xi<1$,
the VEVs of $Z_i^-$ are less than the
cutoff scale. 
This can lead to the FN mechanism.

At this stage, among the fields $Z_i^-$, we define the FN field 
$\Theta$ as the field 
whose VEV mainly compensates for the FI parameter $\xi$.
If we fix the normalization of $U(1)_A$ charge  so that
$\theta=-1$,
then from $D_A=0$, the VEV of the FN field $\Theta$ is determined as 
\begin{equation}
\VEV{\Theta}\equiv\lambda\sim \xi,
\label{lambda}
\end{equation}
which breaks $U(1)_A$ gauge symmetry. 
The other VEVs are determined by the $F$-flatness conditions with 
respect to  $Z_i^+$ as
$\VEV{Z_i^-}\sim \lambda^{-z_i^-}$, which is shown below.
Since $\VEV{Z_i^+}=0$, it is sufficient to examine the terms linear
in $Z_i^+$ in the superpotential in order to determine 
$\VEV{Z_i^-}$. Therefore, in general, 
the superpotential to determine the VEVs can be written
\begin{eqnarray}
W&=&\sum_i^{n_+}W_{Z_i^+}, \label{W}\\
W_{Z_i^+}&=& \lambda^{z_i^+}Z_i^+\left(1+\sum_j^{n_-}\lambda^{z_j^-}Z_j^-
+\sum_{j,k}^{n_-}\lambda^{z_j^-+z_k^-}Z_j^-Z_k^-+\cdots \right) \nn \\
&=&\tilde Z_i^+\left(1+\sum_j^{n_-}\tilde Z_j^-
+\sum_{j,k}^{n_-}\tilde Z_j^-\tilde Z_k^-+\cdots \right), \label{superpot}
\end{eqnarray}
where $\tilde Z_i\equiv \lambda^{z_i}Z_i$.\footnote{
The introduction of discrete symmetries or rational number charges
disallows some of the interactions in Eq. (\ref{superpot}).
However, the results we obtain do not change unless the
situation discussed in the next footnote is realized.}
The $F$-flatness conditions of the $Z_i^+$ fields require
\begin{equation}
\lambda^{z_i^+}\left(1+\sum_j\tilde Z_j^-+\cdots\right)=0,
\end{equation}
which generally lead to solutions $\tilde Z_j^-\sim O(1)$
if these $F$-flatness conditions determine the VEVs.
Thus the F-flatness condition requires
\begin{equation}
   \VEV{ Z_j^-} \sim O(\lambda ^{-z_j^-}).
\label{VEV}
\end{equation}
Note that if there is another field $Z_i^-$ which has larger charge
than the FN field $\Theta$, then the VEV of $Z_i^-$ becomes
larger than $\xi$. This is inconsistent with $D_A=0$.
Therefore it is natural that the field with the 
largest negative
charge becomes the FN field.
Note that if $n_+=n_-$, generically all the VEVs of $Z_i^-$ are
fixed, and therefore there is no flat direction in the potential.
Hence in this case there is no massless field.
Contrasting, if $n_+<n_-$, generally the $n_+$ equations for the
$F$-flatness and $D$-flatness conditions do not determine all the
VEVs of $n_-$ fields $Z_i^-$. Therefore, there are flat directions
in the potential, producing some massless fields.
Thus, if we want to realize the case with no massless mode
in the Higgs sector, $n_+=n_-$ must be imposed in the Higgs
sector.\footnote{
These rough arguments regarding the order of the VEVs and of the 
number counting are based on the assumption 
that the Higgs sector has no other structure by which the difference
between the number of non-trivial $F$-flatness conditions and 
the degree of freedom of non-vanishing VEVs is changed. 
Such a structure can be realized 
 by imposing a certain symmetry, for example, $Z_2$ parity 
(or R parity), or by introducing half integer (or rational number) 
charges.
 When the number of the negatively charged odd $Z_2$-parity fields is
 different from that of positively charged odd $Z_2$-parity fields,
 choosing vanishing VEVs of odd $Z_2$-parity fields changes the difference.
}

If Vacuum A is selected, the anomalous $U(1)_A$ gauge symmetry
is broken at the cutoff scale, and the FN mechanism does not act.
Therefore, we cannot surmise the existence of the $U(1)_A$ gauge 
symmetry from the low energy physics. However, if the Vacuum B is
selected, the FN mechanism acts effectively, and 
the signature of the $U(1)_A$ gauge symmetry can be observed in 
the low energy physics. 
Therefore, it is reasonable to assume that Vacuum B is
selected in our scenario, in which the $U(1)_A$ gauge symmetry
plays an important role in the FN mechanism. The VEVs of
the fields $Z_i^+$ vanish. This guarantees that the SUSY zero 
mechanism\footnote{
Note that if the total charge
of an operator is negative, the $U(1)_A$ invariance and 
analytic property of the superpotential forbids 
the existence of the operator in the superpotential, 
since the field $\Theta$ with negative 
charge cannot compensate for the negative total charge of the operator 
(SUSY zero mechanism). } acts effectively.

To this point, we have examined the VEVs of only singlet fields.
We now consider the case in which there are non-trivial 
representations of the gauge group. 
The same arguments can be applied if we use a set of independent gauge 
invariant operators 
instead of the gauge singlet fields $Z_i$. The gauge invariant
operator $O$ with negative charge $o$ has
non-vanishing VEV $\VEV{O}\sim \lambda^{-o}$ if the $F$-flatness
conditions determine the VEV. For example, let us introduce 
the fundamental representation $C({\bf 27})$ and 
$\bar C({\bf \overline{27}})$ of $E_6$. 
The VEV of the gauge singlet operator $\bar CC$ is estimated as
$\VEV{\bar CC}\sim \lambda^{-(c+\bar c)}$. The essential difference
appears in the $D$-flatness condition
of $E_6$ gauge theory, which requires 
\begin{equation} 
\left|\VEV{C}\right|=\left|\VEV{\bar C}\right|\sim \lambda^{-(c+\bar c)/2}.
\end{equation}
Note that these VEVs are also determined by the anomalous $U(1)_A$
charges, but they are different from the naive expectation
$\VEV{C}\sim\lambda^{-c}$.
This is because the $D$-flatness condition strongly constraints 
the VEVs of non-singlet fields.
One more important $D$-flatness condition is
\begin{equation}
D_A=g_A\left(\xi^2+\sum_i\phi_i|\Phi_i|^2\right)=0.
\end{equation}
The argument for the singlet fields cannot be applied directly to 
the case of the non-trivial representation fields $\Phi_i$.
For example, if we adopt the fields $\Phi({\bf 27})$ and 
$\bar \Phi({\bf \overline{27}})$ of $E_6$ with the charges
 $\phi+\bar \phi=-1$, then the above $D$-flatness condition of
 the anomalous $U(1)_A$ gauge symmetry requires
$\xi^2+\phi|\Phi|^2+\bar \phi |\bar \Phi|^2=0$. The $D$-flatness 
condition of the $E_6$ gauge group leads to $|\Phi|=|\bar \Phi|$, 
and therefore we obtain $\xi^2+(\phi+\bar \phi)|\Phi|^2=0$, which 
implies $|\VEV{\Phi}|=\left|\VEV{\bar \Phi}\right|=\xi$. 
In this case, since $\bar \Phi\Phi$ plays the same role as $\Theta$,
the unit of the hierarchy becomes $\VEV{\bar \Phi\Phi}=\lambda\sim \xi^2$, 
which
is different from the previous relation (\ref{lambda}).
This implies that even if $\xi$ has a 
milder hierarchy, the unit of the hierarchy becomes stronger.

Of course we can determine the VEVs of the singlet operators
from the same superpotential as (\ref{W}),
replacing the singlet fields $Z_i$ by a set of independent gauge singlet 
operators. This, however, is not easy.
However, the situation can be simplified if all the fields 
$\Phi_i^+$~(including non-singlets) 
with positive charges have
vanishing VEVs.
We can obtain the superpotential to determine the VEVs as
\begin{equation}
W=\sum_i^{n_+}W_{\Phi_i^+},
\end{equation}
where $W_X$ denotes the terms linear in the $X$ field.
Note that generally fields with positive charges can have 
non-vanishing VEVs, even if all the gauge singlet operators 
with positive charges have vanishing VEVs. For example, 
if we set $\phi=-3$ and $\bar \phi=2$, then the gauge singlet 
operator $\bar \Phi\Phi$ can have non-vanishing VEV. 
This means that $\bar \Phi$ with positive charge $\bar \phi=2$ has 
a non-vanishing VEV. In such cases, it is not guaranteed that 
the $F$-flatness conditions of fields with negative charges are
automatically satisfied. Therefore we have to take account of the 
superpotential $W(\bar \Phi)$, which includes positively charged fields 
$\bar \Phi$ with non-vanishing VEVs.
We consider such examples below.

In summary, we have the following: 
\begin{enumerate}
\item
Gauge singlet operators with positive total charge have vanishing VEVs,
in order for the FN mechanism to act effectively. This guarantees that 
the SUSY zero mechanism acts effectively.
\item
The singlet operator $\Theta$ with the largest negative charge 
becomes the FN field. When the singlet operator is just 
a singlet field, the VEV is given by $\VEV{\Theta}\sim \xi$, 
which is determined from $D_A=0$. When the singlet operator 
is a composite operator, $\Theta\sim \bar \Phi\Phi$, the VEV 
is given by $\VEV{\Theta}=\xi^2$.
\item
The $F$-flatness conditions of singlet operators with positive 
charges determine the 
VEVs of singlet operators $O$ with negative charges $o$ as
$\VEV{O}\sim \lambda^{-o}$, while the $F$-flatness conditions 
of the singlet operators with negative charges are automatically 
satisfied.
When the operator is a composite operator, $O\sim \bar CC$, the
$D$-flatness condition requires 
$\left|\VEV{C}\right|=\left|\VEV{\bar C}\right|\sim 
\lambda^{-(c+\bar c)/2}$.
\item
If the number of the independent singlet operators~(moduli) with 
positive charges equals that of the fields with negative charges, 
generically no massless fields appear.
\item
The general superpotential to determine the VEVs is expressed 
as $W=\sum_i W_{O_i^+}$, where $W_{O_i^+}$ is linear in the independent
singlet operator $O_i^+$ with positive charges.
When all the fields $\Phi^+$ (including non-singlets) with 
positive charges have vanishing VEVs, the superpotential can 
be written $W=\sum_i^{n_+}W_{\Phi_i^+}$. If some of the 
positively charged fields $\bar \Phi$ have non-vanishing VEVs, 
the superpotential $W_{NV}$ must be added, which includes
only the fields with non-vanishing VEVs.
\end{enumerate}

\section{Doublet-triplet splitting in $SO(10)$ GUT}
Here we make a quick review of the $SO(10)$  unified scenario 
proposed by one of the present authors
\cite{maekawa,maekawa3}.

\subsection{Alignment and DT splitting}
The content of the Higgs sector in $SO(10)\times U(1)_A$ is 
listed in Table I. 
\begin{table}
\begin{center}
Table I. Typical values of anomalous $U(1)_A$ charges.\\
\vspace{1mm}
\begin{tabular}{|c|c|c|} 
\hline
                  &   non-vanishing VEV  & vanishing VEV \\
\hline 
{\bf 45}          &   $A(a=-1,-)$        & $A'(a'=3,-)$      \\
{\bf 16}          &   $C(c=-3,+)$        
                  & $C'(c'=2,-)$      \\
${\bf \overline{16}}$&$\bar C(\bar c=0,+)$ 
                  & $\bar C'(\bar c'=5,-)$ \\
{\bf 10}          &   $H(h=-3,+)$        & $H'(h'=4,-)$      \\
{\bf 1}           &$\Theta(\theta=-1,+)$,$Z(z=-2,-)$,
                  $\bar Z(\bar z=-2,-)$& $S(s=3,+)$ \\
\hline
\end{tabular}
\end{center}
\end{table}
Here the symbols $\pm$ denote the $Z_2$ parity.
The adjoint Higgs
field $A$, whose VEV 
$\VEV{A({\bf 45})}_{B-L}=\tau_2\times {\rm diag}
(v,v,v,0,0)$, breaks $SO(10)$ into
$SU(3)_C\times SU(2)_L\times SU(2)_R\times U(1)_{B-L}$.
 This Dimopoulos-Wilczek form of the VEV plays an
important role in solving the DT splitting problem.
The spinor Higgs fields 
$C$ and $\bar C$ break $SU(2)_R\times U(1)_{B-L}$ into $U(1)_Y$ 
by developing $\VEV{C}(=\VEV{\bar C}=\lambda^{-(c+\bar c)/2}$).
The  Higgs field $H$ contains the usual $SU(2)_L$ doublet.
The gauge singlet operators $A^2$, $\bar CC$ and $H^2$ must
have negative total anomalous $U(1)_A$ charges
to obtain non-vanishing VEVs, 
as discussed 
in the previous section.
Then, in order to give masses to all the Higgs fields,
we have introduced the same number of fields with positive charges,
\footnote{
Strictly speaking, since some of the Higgs fields are absorbed by the 
Higgs mechanism, in principle, a smaller number of positive fields 
can give superheavy masses
to all the Higgs fields. Here we do not examine the possibilities.}
which we denote $A^\prime$, $C^\prime$, $\bar C^\prime$ and
$H^\prime$.
This is, in a sense, a minimal set of the Higgs content.

Following the general argument of the previous section,
the superpotential required by determination of the VEVs can be written
\begin{equation}
W=W_{H^\prime}+ W_{A^\prime} + W_S + W_{C^\prime}+W_{\bar C^\prime}
+W_{NV}.
\end{equation}
Here $W_X$ denotes the terms linear in the positive charged field
$X$, which has vanishing VEV.  Note, however, that terms 
including two
fields with vanishing VEVs like 
$\lambda^{2h^\prime}H^\prime H^\prime$
give contributions to the mass terms but not to the VEVs.
All the terms in $W_{NV}$ contain only the fields with non-vanishing 
VEVs. In the typical charge assignment, 
it is easily checked that they do not play a significant role in our 
argument, since they
do not include the products of only the neutral components under the
standard gauge group. In the following argument, for simplicity, we
ignore the terms that do not include the products of only
neutral components under the standard gauge group, like
${\bf 16}^4$, ${\bf \overline{16}}^4$, ${\bf 10\cdot 16}^2$,
 ${\bf 10 \cdot \overline{16}}^2$ and ${\bf 1\cdot 10}^2$, 
 even if these terms
are allowed by the symmetry.\footnote{ 
 It is easy to include these terms in
our analysis. They can introduce some constraints 
on the vacua other than the standard vacuum, but not 
on the standard vacuum.
}

We now discuss the determination of the VEVs.
If $-3a\leq a^\prime < -5a$,
the superpotential $W_{A^\prime}$ is in general
written 
\begin{equation}
W_{A^\prime}=\lambda^{a^\prime+a}\alpha A^\prime A+\lambda^{a^\prime+3a}(
\beta(A^\prime A)_{\bf 1}(A^2)_{\bf 1}
+\gamma(A^\prime A)_{\bf 54}(A^2)_{\bf 54}),
\end{equation}
where the suffices {\bf 1} and {\bf 54} indicate the representation 
of the composite
operators under the $SO(10)$ gauge symmetry, and $\alpha$, 
$\beta$ and $\gamma$ are parameters of order 1. Here we assume 
$a+a^\prime+c+\bar c<0$
to forbid the term $\bar C A^\prime A C$, which destabilizes the 
DW form of the VEV $\VEV{A}$. 
The $D$-flatness condition requires the VEV
$\VEV{A}=\tau_2\times {\rm diag}(x_1,x_2,x_3,x_4,x_5)$, and the 
$F$-flatness conditions of the $A^\prime$ field requires
$x_i(\alpha\lambda^{-2a}
+(2\beta-\frac{\gamma}{5})(\sum_j x_j^2)+\gamma x_i^2)=0$. 
This allows only two solutions, $x_i^2=0$ and 
$x_i^2=-\frac{\alpha}{(1-\frac{N}{5})\gamma+2N\beta}\lambda^{-2a}$. 
Here $N=0$ -- 5 is the number of $x_i \neq 0$ solutions.
The DW form is obtained when $N=3$.
Note that the higher terms $A^\prime A^{2L+1}$ $(L>1)$ are 
forbidden by the SUSY zero mechanism. If they were allowed, 
the number of possible VEVs other than the DW form would 
become larger, and thus it would become less natural to obtain 
the DW form. 
This is a critical point of this mechanism, and the anomalous 
$U(1)_A$ gauge symmetry plays an essential role in forbidding 
the undesired terms.
In this way, the scale of the VEV is automatically
determined by the anomalous $U(1)_A$ charge of $A$, as noted in the 
previous section.

Next, we discuss the $F$-flatness condition of $S$, which determines
the scale of the VEV $\VEV{\bar C C}$. 
$W_S$ is given by
\begin{equation}
W_S=\lambda^{s+c+\bar c}S\left((\bar CC)+\lambda^{-(c+\bar c)}
+\sum_k\lambda^{-(c+\bar c)+2ka}A^{2k}\right)
\end{equation}
if 
$s\geq -(c+\bar c)$.
Then, the $F$-flatness condition of $S$ implies $\VEV{\bar CC}\sim 
\lambda^{-(c+\bar c)}$, and the $D$-flatness condition requires 
$|\VEV{C}|=|\VEV{\bar C}|\sim \lambda^{-(c+\bar c)/2}$.
The scale of the VEV is again determined only by the charges of 
$C$ and $\bar C$.
If we set $c+\bar c=-3$, then we obtain the VEVs of the fields 
$C$ and $\bar C$
as 
$\lambda^{3/2}$, which differ from the expected values $\lambda^{-c}$ 
and
$\lambda^{-\bar c}$ in the case $c\neq \bar c$.

Next, we discuss the $F$-flatness of $C^\prime$ and 
$\bar C^\prime$,
which causes the alignment of the VEVs $\VEV{C}$ and $\VEV{\bar C}$
and imparts masses on the PNG fields. 
This simple mechanism
was proposed by Barr and Raby
\cite{BarrRaby}.
We can easily assign anomalous $U(1)_A$ charges that allow the 
following superpotential:
\begin{eqnarray}
W_{C^\prime}&=&
       \bar C(\lambda^{\bar c^\prime +c+a}A
       +\lambda^{\bar c^\prime +c+\bar z}\bar Z)C^\prime, \\
W_{\bar C^\prime}&=&
       \bar C^\prime(\lambda^{\bar c^\prime +c+a} A
       +\lambda^{\bar c^\prime +c+z}Z)C.
\end{eqnarray}
The $F$-flatness conditions $F_{C^\prime}=F_{\bar C^\prime}=0$ give
$(\lambda^{a-z} A+Z)C=\bar C(\lambda^{a-\bar z} A+\bar Z)=0$. 
Recall that the VEV of $A$ is 
proportional to the $B-L$ generator $Q_{B-L}$ 
(precisely, $\VEV{A}=\frac{3}{2}vQ_{B-L}$), and that
 $C$, ${\bf 16}$, is decomposed into 
$({\bf 3},{\bf 2},{\bf 1})_{1/3}$, 
$({\bf \bar 3},{\bf 1},{\bf 2})_{-1/3}$, 
$({\bf 1},{\bf 2},{\bf 1})_{-1}$ and $({\bf 1},{\bf 1},{\bf 2})_{1}$ 
under
$SU(3)_C\times SU(2)_L\times SU(2)_R\times U(1)_{B-L}$.
Since $\VEV{\bar CC}\neq 0$, 
 $Z$ is fixed such that $Z\sim -\frac{3}{2}\lambda v Q_{B-L}^0$, 
where $Q_{B-L}^0$ is
the $B-L$ charge of the component of $C$ that has non-vanishing VEV. 
Once the VEV of $Z$ is determined, 
no other component fields can have non-vanishing VEVs, 
because they have different charges $Q_{B-L}$. 
If  the $({\bf 1},{\bf 1},{\bf 2})_1$ field obtains
a non-zero VEV (and therefore $\VEV{Z}\sim -\frac{3}{2}\lambda v$), 
then the gauge group 
$SU(3)_C\times SU(2)_L\times SU(2)_R\times U(1)_{B-L}$ is broken 
down to the standard gauge group. 
Once the direction of the VEV $\VEV{C}$ is 
determined, the VEV $\VEV{\bar C}$ must have the same direction, 
because of the $D$-flatness
condition. Therefore, $\VEV{\bar Z}\sim -\frac{3}{2}\lambda v$.

Finally the $F$-flatness condition of $H^\prime$ is examined. 
$W_{H^\prime}$ is
written
\begin{equation}
W_{H^\prime}=\lambda^{h+a+h^\prime}H^\prime AH. 
\end{equation}
$F_{H^\prime}$ leads to a vanishing VEV of the color triplet
Higgs, $\VEV{H_T}=0$. 
All VEVs have now been fixed.

There are several terms that must be forbidden for the stability 
of the DW mechanism. For example, $H^2$, $HZH^\prime$ and 
$H\bar Z H^\prime$ induce a large mass of the doublet Higgs, 
and the term $\bar CA^\prime A C$ would destabilize the DW form of 
$\VEV{A}$.
We can easily forbid these terms using the SUSY zero mechanism.
For example, if we choose
$h<0$, then $H^2$ is forbidden, and if we choose $\bar c+c+a+a^\prime<0$, 
then
$\bar CA^\prime A C$ is forbidden. 
Once these dangerous terms are forbidden
by the SUSY zero mechanism, higher-dimensional terms that also become
dangerous (for example, 
$\bar CA^\prime A^3 C$ and $\bar CA^\prime C\bar CA C$) are automatically
forbidden.
This is also an advantageous property of our scenario. 

To end this subsection, we would like to explain how to determine
the symmetry and the quantum numbers in the Higgs sector to realize 
DT splitting.
It is essential that dangerous terms be forbidden by the SUSY zero 
mechanism, and the necessary terms must be allowed by the symmetry. 
The dangerous terms are
\begin{equation}
H^2, HH',HZH', \bar CA'C, \bar CA'AC, \bar CA'ZC, A'A^4, A'A^5.
\end{equation}
The terms required to realize DT splitting are
\begin{equation}
A'A, A'A^3, HAH',\bar C'(A+Z)C, \bar C(A+Z)C', S\bar CC.
\end{equation}
Here we denote both $Z$ and $\bar Z$ as ``$Z$''.
In order to forbid $HH'$ but not $HAH'$, 
we introduce $Z_2$ parity.

Of course, the above conditions are necessary but not sufficient. 
To determine whether a given assignment actually works well,
we have to write down the mass matrices of Higgs sector. 
This is done in the next subsection.

\subsection{Mass spectrum of the Higgs sector}
Under $SO(10)\supset SU(5) \supset SU(3)_C\times SU(2)_L\times U(1)_Y$,
the spinor ${\bf 16}$, vector ${\bf 10}$ and the adjoint ${\bf 45}$
are classified in terms of the fields  $Q({\bf 3,2})_{\frac{1}{6}}$,
$U^c({\bf \bar 3,1})_{-\frac{2}{3}}$, $D^c({\bf \bar 3,1})_{\frac{1}{3}}$,
$L({\bf 1,2})_{-\frac{1}{2}}$, $E^c({\bf 1,1})_1,N^c({\bf 1,1})_0$,
$X({\bf 3,2})_{-\frac{5}{6}}$ and their conjugate fields, and
$G({\bf 8,1})_0$ and $W({\bf 1,3})_0$ as
\begin{eqnarray}
{\bf 16}&= &
\underbrace{[Q+U^c+E^c]}_{\bf 10}+\underbrace{[D^c+L]}_{\bf \bar 5}
+\underbrace{N^c}_{\bf 1}, \nn \\
{\bf 10}&= &
\underbrace{[D^c+L]}_{\bf \bar 5}+\underbrace{[\bar D^c+\bar L]}_{\bf 5},
\label{class}\\
{\bf 45}&= &
\underbrace{[G+W+X+\bar X+N^c]}_{\bf 24}
+\underbrace{[Q+U^c+E^c]}_{\bf 10}
+\underbrace{[\bar Q+\bar U^c+\bar E^c]}_{\bf \overline{10}}
+\underbrace{N^c}_{\bf 1}. \nn
\end{eqnarray}

In the following, we study how mass matrices of the above fields
are determined by considering an example of the typical charge 
assignment given in Table I.
For the mass terms, we must
take account of not only the terms in the previous section but also
the terms that contain two fields with vanishing
VEVs. 

First, we examine the mass spectrum of ${\bf 5}$ and ${\bf \bar 5}$ of
$SU(5)$. 
Considering the additional terms
$\lambda^{2h^\prime} H^\prime H^\prime$,
$\lambda^{c^\prime+\bar c^\prime}\bar C^\prime C^\prime$, 
$\lambda^{c'+c+h'}C'CH'$, 
$\lambda^{\bar c'+z+\bar c+h}Z\bar C'\bar CH$,
$\lambda^{\bar c'+\bar c+h'}\bar C'\bar CH'$ and
$\lambda^{2\bar c+h'}\bar C^2H'$,
the mass matrices $M_I$ ($I=D^c(H_T),L(H_D)$),
whose elements correspond to the mass of the $I$ component of
${\bf \bar 5(10)}$ or ${\bf \bar 5(16)}$ and 
the $\bar I$ component of ${\bf 5(10)}$ or ${\bf 5(\overline{16})}$, 
are given as 
\begin{equation}
M_I=\bordermatrix{
\bar I\backslash I&10_H(-3) & 16_{C}(-3) & 
                  10_{H^\prime}(4)&16_{C^\prime}(2) \cr
10_H(-3) & 0 & 0 & \lambda^{h+h^\prime +a}\VEV{A}  & 0 \cr
\overline{16}_{\bar C}(0)& 0 & 0 & \lambda^{h'+2\bar c}\VEV{\bar C} 
& \lambda^{\bar c+c^\prime} \cr
10_{H^\prime} (4)& \lambda^{h+h^\prime +a}\VEV{A} & 0& \lambda^{2h^\prime} 
 & \lambda^{h'+c'+c}\VEV{C} \cr
\overline{16}_{\bar C^\prime}(5) &  
\lambda^{h+\bar c'+\bar c}\VEV{\bar C} 
 & \lambda^{c+\bar c^\prime} 
 & \lambda^{h'+\bar c'+\bar c}\VEV{\bar C}
  & \lambda^{c^\prime+\bar c^\prime} \cr},
\end{equation}
where the vanishing components result from the SUSY zero mechanism,
and we indicate typical charges in parentheses.

It is worthwhile examining the general structure of the 
mass matrices.
The first two columns and rows correspond to the fields with 
non-vanishing VEVs that have smaller charges, and the last 
two columns and rows correspond to the fields with vanishing VEVs 
that have larger charges.
Therefore, it is useful to divide the matrices into four $2\times 2$ 
matrices as
\begin{equation}
M_I=\left(\matrix{ 0 & A_I \cr B_I & C_I }\right).
\end{equation}
It is easily seen 
that the ranks of $A_L$ and $B_L$ are reduced to 1
when the VEV $\VEV{A}$ vanishes.
This implies that
the rank of $M_L$ is reduced, and actually it becomes 3.
However, the ranks of $A_{D^c}$ and $B_{D^c}$ remain 2, 
because the field $A$ becomes non-zero on $D^c$.
Therefore DT splitting is realized.
The mass spectrum of $L$ is easily obtained as
$(0,\lambda^{2h'},\lambda^{\bar c+c'},\lambda^{\bar c'+c})$.
The massless modes of the doublet Higgs are approximately given by
\begin{equation}
5_H, \bar 5_H+\lambda^{h-c+\frac{1}{2}(\bar c-c)}\bar 5_C.
\label{mixing}
\end{equation}
The elements of the 
matrices $A_I$ and $B_I$  become generally larger than the 
elements of the matrices $C_I$ because the total anomalous 
$U(1)_A$ charge of the corresponding pair of fields in $A_I$ 
and $B_I$ becomes smaller than that in $C_I$.
Therefore, the mass spectrum of $D^c$ is essentially determined 
by the matrices $A_I$ and $B_I$ as 
$(\lambda^{h+h'},\lambda^{h+h'},\lambda^{\bar c+c'},
\lambda^{\bar c'+c})$.
It is obvious that to realize proton decay, we have to pick up 
an element of $C_I$. 
Since such an element is generally smaller than the mass scale 
of $D^c$, proton decay is suppressed.
The effective colored Higgs mass is estimated as
$(\lambda^{h+h^\prime})^2/\lambda^{2h^\prime}=\lambda^{2h}$, 
which is larger than the cutoff scale, because
$h<0$.

Next, we examine the mass matrices for the representations 
$I=Q,U^c$ and $E^c$,
which are contained in the {\bf 10} of $SU(5)$,
where the additional terms
$\lambda^{2a^\prime}A^\prime A^\prime$, 
$\lambda^{c^\prime+\bar c^\prime}\bar C^\prime C^\prime$,
$\lambda^{c^\prime+a^\prime+\bar c} \bar CA^\prime C^\prime$ and
$\lambda^{\bar c^\prime+a^\prime+c} \bar C^\prime A^\prime C$
must be taken into account.
The mass matrices are written
\begin{equation}
M_I=\bordermatrix{
\bar I\backslash I&45_A(-1) &16_{ C}(-3)& 45_{A^\prime}(3) &
16_{C^\prime}(2) \cr
45_A(-1) &0& 0 & \lambda^{a^\prime+a} \alpha_I  & 
\lambda^{\bar c+c^\prime+a}\VEV{\bar C} \cr
\overline{16}_{\bar C}(0)&0 & 0 & 0 & \lambda^{\bar c+c^\prime}\beta_I \cr
45_{A^\prime}(3) &\lambda^{a+a^\prime} \alpha_I & 0& \lambda^{2a^\prime}  & 
\lambda^{\bar c+c^\prime+a^\prime}\VEV{\bar C} \cr
\overline{16}_{\bar C^\prime}(5) &\lambda^{c+\bar c^\prime+a}
\VEV{C} &\lambda^{c+\bar c^\prime}\beta_I &
\lambda^{c+\bar c^\prime+a^\prime}\VEV{C} &
 \lambda^{c^\prime+\bar c^\prime}\cr},
\label{mass10}
\end{equation}
where $\alpha_Q=\alpha_{U^c}=0$ and $\beta_{E^c}=0$, because
there are NG modes in symmetry breaking processes
$SO(10)\rightarrow SU(3)_C\times SU(2)_L\times SU(2)_R\times U(1)_{B-L}$
and $SU(2)_R\times U(1)_{B-L}\rightarrow U(1)_Y$, respectively.
Defining $2\times 2$ matrices as in the $I=L,D^c$ case,
it is obvious that the ranks of $A_I$ and $B_I$ are reduced.
Thus for each $I$, the $4\times 4$ matrices $M_I$ have one 
vanishing eigenvalue, which corresponds to the NG mode absorbed
by the Higgs mechanism. The mass spectrum of the remaining three
modes is ($\lambda^{c+\bar c^\prime}$, $\lambda^{c^\prime+\bar c}$,
$\lambda^{2a^\prime}$) for the color-triplet modes $Q$ and $U^c$, and
($\lambda^{a+a^\prime}$, 
$\lambda^{a+a^\prime}$,
$\lambda^{c^\prime+\bar c^\prime}$) or 
($\lambda^{c+\bar c^\prime+a}\VEV{C}$, 
$\lambda^{c^\prime+\bar c+a}\VEV{\bar C}$,
$\lambda^{2a^\prime}$) for the color-singlet modes $E^c$.

The adjoint fields $A$ and $A^\prime$ contain
two $G$, two $W$ and two pairs of $X$ and $\bar X$, 
whose mass matrices $M_I(I=G,W,X)$ are given by
\begin{equation}
M_I=\bordermatrix{
\bar I\backslash I &    45_A(-1)       &        45_{A'}(3)           \cr
45_A(-1)    &     0        & \alpha_I\lambda^{a+a'}  \cr
45_{A'}(3) & \alpha_I\lambda^{a+a'} & \lambda^{2a'}  \cr}.
\end{equation}
Two $G$ and two $W$ acquire masses $\lambda^{a^\prime+a}$.
Since $\alpha_X=0$, one pair of $X$ is massless, and this is absorbed
by the Higgs mechanism. 
The other pair has a rather light mass of $\lambda^{2a^\prime}$.

\subsection{Gauge unification and proton decay}
In the minimal SUSY $SU(5)$ GUT, proton stability is not compatible
with the success of the gauge coupling unification
\cite{Goto}.
Proton stability requires the colored Higgs mass to be 
larger than $10^{18}$ GeV,
\footnote{
In Ref. \cite{Goto}, the lower bound of the colored Higgs mass was
obtained as $6.5\times 10^{16}$ GeV. To derive this value, the 
hadron matrix element parameter $\alpha=0.003$ and $\tan \beta=2.5$
were used.
If we use the result of a recent lattice calculation, 
$\alpha\sim 0.015$ \cite{lattice},
and a more reasonable (larger) value of $\tan \beta$, the lower bound 
easily become larger. } 
which destroys the coupling unification, 
because it has no other tuning parameter. Of course, if we introduce
other superheavy particles with masses smaller than the GUT scale,
we may recover the gauge coupling unification by tuning their masses. 
However, unless we
have some mechanism that controls the scale of the masses, generally
some fine-tuning is required. 
In the $SO(10)$ GUT with the DW mechanism, proton stability can be
realized in the mass structure
\begin{equation}
M_I=\left(\matrix{0 & \VEV{A} \cr \VEV{A} & m \cr}\right) (I=L, D_c),
\end{equation}
if $\VEV{A}^2/m_{D^c}>10^{18}$ GeV. However, in order to realize the 
condition $\VEV{A}^2/m_{D^c}>10^{18}$ GeV, the mass scale of 
the additional doublet Higgs $m$ becomes smaller than the GUT scale
$\VEV{A}$, which generally destroys the success of the gauge coupling
unification. Of course, it may be possible to realize 
gauge coupling unification by tuning the other scale, $\VEV{C}$,
or the mass scales of superheavy particles.
Unless we have no mechanism that controls these scales, however, 
such a situation cannot explain why the gauge couplings meet
at a scale $\Lambda_G\sim 2\times 10^{16}$ GeV in the minimal SUSY
standard model (MSSM).

In our scenario, 
once we determine the anomalous $U(1)_A$ charges, the mass spectra
of all superheavy particles and other symmetry breaking scales are
 determined, and hence we can examine whether or not 
the running couplings from the low energy scale meet at the 
unification scale. When the initial values of the gauge coupling 
constants are replaced by the usual GUT scale, 
$\Lambda_G\sim 2\times 10^{16}$ GeV, and the unified gauge coupling, 
the condition for gauge coupling unification can be converted into 
a relation between the charges and the ratio of the cutoff scale 
$\Lambda$ to the usual GUT scale $\Lambda_G\sim 2\times 10^{16}$ GeV:
\begin{equation}
\frac{\Lambda_G}{\Lambda}\sim \lambda^{-\frac{h}{7}}
                         \sim \lambda^{\frac{h}{8}}.
\end{equation}
This leads to 
\begin{eqnarray}
\Lambda&\sim& \Lambda_G, \\
h&\sim &0.
\end{eqnarray}
Here we have used the renormalization group up to the one loop 
approximation.
It is non-trivial that
in this relation, all the charges except that of the Higgs doublet 
are cancelled out. 
If we simply had taken $\Lambda=\Lambda_G$ and $h=0$, the model 
would exhibit proton decay via dimension 5 operators, 
because the effective colored Higgs mass becomes 
$\lambda^{2h}\Lambda=\Lambda_G\ll10^{18}$ GeV. 
However, we have to use a negative $h$ to forbid the Higgs mass term
$H^2$. Therefore, we would like to know how large a negative charge
$h$ can be adopted in our scenario.
To obtain realistic quark and lepton mass matrices
including bi-large neutrino mixing, the maximal value of $h$ is $-3$.
In this case, the effective colored Higgs mass becomes
$\lambda^{2h}\Lambda=\lambda^{-6}\Lambda_G\sim 10^{22}$ GeV, which
is much larger than the present experimental limit. 
Note that even for such a small value of $h$, 
coupling unification can be realized,
using the ambiguities of order 1 coefficients. 
Since the unified scale becomes $\lambda^{-a}\Lambda$, 
just below the scale $\Lambda_G$, we believe that proton decay 
via dimension 6 operators will be observed in the near future.
We will return to this point in the next section.

Once we have fixed the anomalous $U(1)_A$ charges, the fact that the 
gauge couplings meet at the usual GUT scale in the MSSM is 
non-trivially related to the 
result that the gauge couplings of the GUT with anomalous $U(1)_A$ 
gauge symmetry  almost meet at the GUT scale $\lambda^{-a}\Lambda$ 
in our scenario. Therefore, this GUT scenario can explain why the 
gauge couplings meet at a scale in MSSM with an accuracy up to the 
one loop approximation.%
\footnote{ Actually,
if it were the case that the gauge couplings meet at the other scale 
$\Lambda_O$ in MSSM, 
then the cutoff scale 
would be the scale $\Lambda_O$ in our scenario; that is, the GUT scale 
would be $\lambda^{-a}\Lambda_O$. }

\section{$E_6$ unification of the Higgs sector}
In this section, we extend the DT splitting mechanism, discussed in the
previous section, to $E_6$ unification. Here we propose the complete
Higgs sector with the $E_6$ GUT gauge group.

In order to break the $E_6$ gauge group into the standard gauge group,
we introduce the following Higgs content:
\begin{enumerate}
\item Higgs fields that break $E_6$ into $SO(10)$:
$\Phi({\bf 27})$ and $\bar \Phi ({\bf \overline{27}})$
($\left|\VEV{\Phi({\bf 1,1})}\right|=
\left|\VEV{\bar \Phi({\bf 1,1})}\right|$).
\item An adjoint Higgs field that breaks $SO(10)$
into 
$SU(3)_C\times SU(2)_L\times SU(2)_R\times U(1)_{B-L}$: $A({\bf 78})$
($\VEV{{\bf 45}_A}=\tau_2\times {\rm diag}(v,v,v,0,0)$).
\item Higgs fields that break $SU(2)_R\times U(1)_{B-L}$ into $U(1)_Y$:
$C({\bf 27})$ and $\bar C({\bf \overline{27}})$
($\left|\VEV{C({\bf 16,1})}\right|=
\left|\VEV{\bar C({\bf \overline{16},1})}\right|$).
\end{enumerate}
Of course, 
 the anomalous $U(1)_A$ charges of 
the gauge singlet operators, $\bar \Phi\Phi$, $\bar CC$ and $A^2$, 
must be negative.

Naively thinking, it appears that we have to introduce at least 
the same number of superfields with positive charges in order to 
make them massive. 
In fact, however, we find this is not the case.
This is because some of the Higgs
fields with non-vanishing VEVs are absorbed by the Higgs mechanism.
Actually, when the $E_6$ gauge group is broken into $SO(10)$ 
by non-vanishing VEV 
$\left|\VEV{\Phi}\right|=\left|\VEV{\bar \Phi}\right|$, the fields
${\bf 16}_\Phi$ and ${\bf \overline{16}}_{\bar\Phi}$ are absorbed 
by the super-Higgs mechanism.
\footnote{
Strictly speaking, a linear combination of $\Phi$, $C$ and $A$
and of $\bar\Phi$, $\bar C$ and $A$
become massive through the super-Higgs mechanism. The main modes are
${\bf 16}_\Phi$ and ${\bf \overline{16}}_{\bar \Phi}$, respectively.
}
Therefore, if two additional {\bf 10}s of $SO(10)$ in the Higgs 
content with non-vanishing VEVs can be massive, then
we can save the superfields with positive charges.
At first glance, such a mass term seems to be forbidden by the 
SUSY zero mechanism.
Actually, if all fields with non-vanishing VEVs had negative 
anomalous $U(1)_A$ charges, their mass term would be forbidden. 
As discussed in \S2, the non-positiveness of the anomalous $U(1)_A$ 
charges is required only for gauge singlet operators with 
non-vanishing VEVs, so even fields with positive charges can 
have non-vanishing VEVs if the total charge of the gauge singlet 
operators with non-vanishing VEVs is negative. 
For example, we can set $\phi=-3$ and $\bar \phi=2$, because
the total charge of the gauge singlet operator $\bar \Phi\Phi$
is negative. Since $\bar \Phi$
has positive charge, the term $\bar \Phi^3$ is allowed, and it 
induces a mass of ${\bf 10}_{\bar \Phi}$ through the non-vanishing 
VEV $\VEV{\bar \Phi}$. If the term $\bar \Phi^2\bar C$ is allowed,
masses of the two ${\bf 10}$s,  ${\bf 10}_{\bar \Phi}$ and 
${\bf 10}_{\bar C}$, are induced, so we can save the superfields
with positive charges.

The minimal content of the Higgs sector with $E_6\times U(1)_A$ gauge 
symmetry is given in Table II, where the symbols $\pm$ denote the 
$Z_2$ parity quantum numbers.
\begin{table}
\begin{center}
Table II. Typical values of anomalous $U(1)_A$ charges.
\\ \vspace{0.1cm}
\begin{tabular}{|c|c|c|} 
\hline
                  &   non-vanishing VEV  & vanishing VEV \\
\hline 
{\bf 78}          &   $A(a=-1,-)$        & $A'(a'=4,-)$      \\
{\bf 27}          &   $\Phi(\phi=-3,+)$\  $C(c=-6,+)$ &  $C'(c'=7,-)$  \\
${\bf \overline{27}}$ & $\bar \Phi(\bar \phi=2,+)$ \  $\bar C(\bar c=-2,+)$ &

                  $\bar C'(\bar c'=8,-)$ \\
{\bf 1}           &   $Z_2(z_2=-2,-)$,$Z_5(z_5=-5,-)$,
                       $\bar Z_5(\bar z_5=-5,-)$ \   &  \\
\hline
\end{tabular}
\end{center}
\end{table}
Here the Higgs field $H$ of the $SO(10)$ model is contained in $\Phi$. 
This $E_6$
Higgs sector has the same number of superfields with non-trivial 
representation as the $SO(10)$ Higgs sector, in spite of the fact
that the larger group $E_6$ requires additional Higgs fields
to break $E_6$ into the $SO(10)$ gauge group.
It is interesting that the DT splitting
is naturally realized in this minimal Higgs content in a sense.

\subsection{DT splitting and alignment}
Generally in $E_6$ GUT, the interactions in the superpotential
of ${\bf 27}$ and ${\bf \overline{27}}$ are written 
in terms of the units ${\bf 27}^3$, ${\bf \overline{27}27}$ and 
${\bf\overline{27}}^3$.
Note that terms like ${\bf 27}^3$ or ${\bf\overline{27}}^3$  
do not contain the product of singlet components of the standard 
gauge group. Therefore these terms can be ignored when considering 
the standard vacuum. 
Of course, these terms can constrain vacua other than the standard
vacuum. This point is discussed below.

The important terms in the superpotential
to determine the VEVs are 
\begin{equation}
W=W_{A^\prime} + W_{C^\prime}+W_{\bar C^\prime}+W({\bar \Phi}).
\end{equation}
Since we have a positively charged field
$\bar \Phi$
which has non-vanishing VEV, we have to take account of the terms
$W({\bar \Phi})$, which include the field $\bar \Phi$ but not
the fields with vanishing VEVs. Since 
$\bar \Phi\Phi$ and $\bar \Phi C$ have negative total charges,
the superpotential essentially has terms like $\overline{27}^3$.
Therefore, the superpotential $W(\bar \Phi)$ can constrain vacua 
other than the standard vacuum. 

Let us examine the superpotential $W(\bar \Phi)$ to elucidate the 
general vacuum structure in the $E_6$ model.
As discussed in the previous section, 
the composite operator $\bar \Phi\Phi$ with $\phi+\bar \phi=-1$ 
can play the same role as
the FN field. In that case, $\Phi$ and $\bar \Phi$ have
non-vanishing VEVs, and the $E_6$ $D$-flatness condition requires
$\VEV{\Phi}=\VEV{\bar \Phi}$, up to phases. The VEV of $\bar \Phi$ 
(and therefore that of $\Phi$ also, by the $D$-flatness condition) 
can be rotated by the $E_6$ gauge transformation into the following 
form:
\begin{equation}
\VEV{\bar \Phi}=
\pmatrix{ \bar u \cr
          {\bf 0} \cr
          \bar u_1 \cr
          \bar u_2 \cr
          {\bf 0}        } 
\begin{array}{l}
  \hbox{$\}SO(10)$ singlet (real)} \\
  \hbox{$\}SO(10)$ ${\bf \overline{16}}$}       \\
  \hbox{$\}$the first component of $SO(10)$ {\bf 10}\ (complex)} \\
  \hbox{$\}$the second component of $SO(10)$ {\bf 10}\ (real)} \\
  \hbox{$\}$the third to tenth components of $SO(10)$ {\bf 10}}.
\end{array}
\end{equation}
For simplicity, we adopt a superpotential of the form 
\begin{equation}
W({\bar \Phi})=\bar \Phi^3+\bar \Phi^2\bar C.
\end{equation}
Then, the $F$-flatness conditions of ${\bf 10}_{\bar C}$ and
${\bf 1}_{\bar C}$ lead to
${\bf 1}_{\bar \Phi}{\bf 10}_{\bar \Phi}=0$ and ${\bf 10}_{\bar \Phi}^2=0$.
Thus we are allowed to have either the vacuum
$\bar u\neq 0,\bar u_1=\bar u_2=0$ or the vacuum 
$\bar u=0,\bar u_1=i\bar u_2\neq 0$.
This implies that the non-vanishing of the VEV 
$\VEV{{\bf 1}_{\bar \Phi}}$ requires
the vanishing of the VEV $\VEV{{\bf 10}_{\bar \Phi}}$. Therefore,
in the first vacuum, the $E_6$ gauge group is broken into the 
$SO(10)$ gauge group.
Moreover, 
in this vacuum, ${\bf 10}_{\bar C}$ has vanishing VEV, because of
the $F$-flatness conditions of ${\bf 10}_{\bar \Phi}$.
Interestingly enough, a vacuum alignment occurs naturally.
In the following, for simplicity, we often write $\lambda^n$ in place 
of the operators $(\bar\Phi\Phi)^n$, though these operators are not 
always singlets.

The superpotential $W_{A^\prime}$ is in general
written as
\begin{eqnarray}
W_{A^\prime}&=&\lambda^{a^\prime+a}A^\prime A
+\lambda^{a^\prime+3a}A^\prime A^3
+\lambda^{a'+a+\bar \phi+\phi}\bar \Phi A'A\Phi \nonumber \\ 
&&+\lambda^{a'+3a+\bar \phi+\phi}\bar \Phi A'A^3\Phi, 
\end{eqnarray}
under the condition $-3a+\bar \phi+\phi\leq a^\prime < -5a$.
Here we assume 
$c+\bar c, c+\bar \phi, \bar c+\phi<-(a'+a)$
to forbid the terms $\bar C A^\prime A C$ (which destabilizes the 
DW form of the VEV of $A$), $\bar CA'A\Phi$ and $\bar \Phi A'AC$
(which may lead to undesired vacua in which $\VEV{\bar C}=\VEV{C}=0$).

If $A$ and $(\Phi,\bar \Phi)$ are separated in the superpotential, 
PNG fields appear. 
Since the terms $\bar \Phi A'A\Phi$ and $\bar \Phi A'A^3\Phi$ connect 
$A'$ and $A$ with $\Phi$ and $\bar \Phi$, the PNG obtain non-zero 
masses.
Moreover, these terms realize the alignment between the VEVs
$\left|\VEV{\Phi}\right|=\left|\VEV{\bar \Phi}\right|$ and $\VEV{A}$. 
Note that these terms
are also important to induce the term 
$({\bf 45}_{A'}{\bf 45}_{A})_{\bf 54}({\bf 45}_{A}^2)_{\bf 54}$,
which is not included in the term $A'A^3$,
 because of a cancellation (see Appendix A).
\footnote{We thank T. Kugo for pointing out this cancellation.}
In terms of 
$SO(10)$, which is not broken by the VEV 
$\left|\VEV{\Phi}\right|=\left|\VEV{\bar \Phi}\right|$,
the effective superpotential is given by
\begin{eqnarray}
W_{A'}^{eff}&=&
{\bf 45}_{A'}(1+{\bf 1}_{A}^2+{\bf 45}_{A}^2
+{\bf \overline{16}}_{A}{\bf 16}_{A}){\bf 45}_{A}  \nn \\
&&+{\bf \overline{16}}_{A'}(1+{\bf 1}_{A}^2+{\bf 45}_{A}^2
+{\bf \overline{16}}_{A}{\bf 16}_{A}){\bf 16}_{A} \nn \\
&&+{\bf 16}_{A'}(1+{\bf 1}_{A}^2+{\bf 45}_{A}^2
+{\bf \overline{16}}_{A}{\bf 16}_{A}){\bf \overline{16}}_{A} \\
&&+{\bf 1}_{A'}{\bf 1}_{A}(1+{\bf 1}_{A}^2+{\bf 45}_{A}^2
+{\bf \overline{16}}_{A}{\bf 16}_{A}). \nn
\end{eqnarray}
The $F$-flatness conditions are written
\begin{eqnarray}
\frac{\partial W}{\partial {\bf 45}_{A'}}&=& (1+{\bf 1}_{A}^2
+{\bf 45}_{A}^2+{\bf \overline{16}}_{A}{\bf 16}_{A}){\bf 45}_{A}, 
\label{45}\\
\frac{\partial W}{\partial {\bf \overline{16}}_{A'}}&=&(1+{\bf 1}_{A}^2
+{\bf 45}_{A}^2+{\bf \overline{16}}_{A}{\bf 16}_{A}){\bf 16}_{A}, 
\label{bar16} \\
\frac{\partial W}{\partial {\bf 16}_{A'}}&=& {\bf \overline{16}}_{A}
(1+{\bf 1}_{A}^2+{\bf 45}_{A}^2+{\bf \overline{16}}_{A}{\bf 16}_{A}), 
\label{16}\\
\frac{\partial W}{\partial {\bf 1}_{A'}}&=& {\bf 1}_{A}
(1+{\bf 1}_{A}^2+{\bf 45}_{A}^2+{\bf \overline{16}}_{A}{\bf 1}_{A}).
\label{1}
\end{eqnarray}
These $F$-flatness conditions and the $D$-flatness conditions of
$SO(10)$ determine the VEVs 
$\VEV{{\bf 16}_A}=\VEV{{\bf \overline{16}}_A}=0$. We have two 
possibilities for the VEV of ${\bf 1}_A$, one vacuum with 
$\VEV{{\bf 1}_A}=0$ and another vacuum with $\VEV{{\bf 1}_A}\neq 0$. 
In the latter vacuum, the DW mechanism in $E_6$ GUT does not act, 
because the non-vanishing VEV $\VEV{{\bf 1}_A}$
directly gives the bare mass to the doublet Higgs. 
Therefore, the former vacuum in which $\VEV{{\bf 1}_A}=0$, is 
desirable to realize DT splitting.
Note that if the term $\bar \Phi A'\Phi$ is allowed, the vacuum
$\VEV{{\bf 1}_A}=0$ disappears. This destroys the realization of
DT splitting. Here this term is forbidden by $Z_2$ parity.
As in the $SO(10)$ case, we have several possibilities for the VEV of 
${\bf 45}_A$, one of which is the DW-type of the VEV 
$\VEV{{\bf 45}_A}_{B-L}=i\tau_2\times {\rm diag}(v,v,v,0,0)$,
where $v\sim\lambda^{-a}$.
These VEVs break the $SO(10)$ gauge group
into $SU(3)_C\times SU(2)_L\times SU(2)_R\times U(1)_{B-L}$.

Next, we discuss the $F$-flatness of $C^\prime$ and 
$\bar C^\prime$,
which not only determine the scale of the VEV 
$\VEV{\bar CC}\sim \lambda^{-(c+\bar c)}$ but also realize 
the alignment of the VEVs $\VEV{C}$ and $\VEV{\bar C}$.
For simplicity, we assume that 
$\VEV{{\bf 1}_C}=\VEV{{\bf 1}_{\bar C}}=0$, 
though there may be vacua in which these components have 
non-vanishing VEVs.
Then, since $\VEV{{\bf 10}_C}=\VEV{{\bf 10}_{\bar C}}=0$ 
by the above argument,
only the components ${\bf 16}_C$ and ${\bf \overline{16}}_{\bar C}$
can have non-vanishing VEVs.
The superpotential to determine these VEVs 
can be written 
\begin{eqnarray}
W_{C^\prime}&=&
       \lambda^{\bar\phi+c'}\bar \Phi (\lambda^{c+\bar c+a}\bar CAC+
       \lambda^{z_5}Z_5+\lambda^{\bar z_5}\bar Z_5+\lambda^{z_2}Z_2+
       \lambda^{a}A)C'
       \nn\\
       &&+ \lambda^{\bar c+c'}\bar C(\lambda^{z_5}Z_5
       +\lambda^{\bar z_5}\bar Z_5
       +\lambda^{z_2}Z_2+
       \lambda^aA)C', \\
W_{\bar C^\prime}&=&
       \lambda^{\bar c'+\phi}\bar C'(\lambda^{z_5}Z_5
       +\lambda^{\bar z_5}\bar Z_5+
       \lambda^{z_2}Z_2+\lambda^aA)\Phi \nn\\
       &&+\lambda^{\bar c'+c}\bar C'(\lambda^{z_2}Z_2+\lambda^aA)C,
\end{eqnarray}
where we omit the even $Z_2$ parity operators with non-vanishing VEVs,
like $A^{2n}$, $Z_2^2$, $Z_iA$, etc., because the VEVs of these 
operators do not change the power of $\lambda$.
The $F$-flatness conditions of
${\bf 1}_{C'}$ and ${\bf 1}_{\bar C'}$ lead to 
$\bar \Phi (\lambda^{c+\bar c+a}\bar CAC+
       \lambda^{z_5}Z_5+\lambda^{\bar z_5}\bar Z_5+
       \lambda^{z_2}Z_2+\lambda^aA)=0$ 
and 
$(\lambda^{z_5}Z_5+\lambda^{\bar z_5}\bar Z_5+
       \lambda^{z_2}Z_2+\lambda^aA)\Phi=0$, 
respectively. The vacua are  
(a) $\VEV{\bar CC}=0$ and  (b) $\VEV{\bar CC}\neq 0$.
The desired vacuum (a) requires the additional $F$-flatness conditions 
of ${\bf 16}_{C'}$ and ${\bf \overline{16}}_{\bar C'}$, which causes
the alignment of the VEVs $\VEV{A}$ and $\VEV{C}(\VEV{\bar C})$, as in 
the $SO(10)$ cases. Then, the above four $F$-flatness conditions with
respect to  ${\bf 1}_{C'}$, ${\bf 1}_{\bar C'}$, ${\bf 16}_{C'}$ and 
${\bf \overline{16}}_{\bar C'}$
determine the scale of
the four VEVs
$\VEV{\bar CC}\sim \lambda^{-(c+\bar c)}$,
$\VEV{Z_i}\sim \lambda^{-z_i}(i=3,5)$ and 
$\VEV{\bar Z_5}\sim \lambda^{-\bar z_5}$.
The VEVs $\left|\VEV{C}\right|=\left|\VEV{\bar C}\right|
\sim \lambda^{-(\bar c+c)}$ 
break $SU(2)_R\times U(1)_{B-L}$ into $U(1)_Y$. 

Thus all the VEVs are determined by the anomalous $U(1)_A$ charges.

\subsection{Mass spectrum of the Higgs sector}
Since all the VEVs are fixed, we can derive the mass spectrum
of the Higgs sector.

Let the fields be decomposed in terms of the quantum numbers
of $SO(10)\times U(1)_{V'}$ as
\begin{eqnarray}
{\bf 27}&=& {\bf 16}_1+{\bf 10}_{-2}+{\bf 1}_4, \\
{\bf 78}&=& {\bf 45}_0+{\bf 16}_{-3}+{\bf \overline{16}}_3+{\bf 1}_0,
\end{eqnarray}
which are further decomposed into $SU(5)$ representations 
[see Eq.~(\ref{class})].

In the following, we study how the mass matrices of the above fields
are determined by anomalous $U(1)_A$ charges.
Note that for the mass terms, we must
take account of not only the terms given in the previous subsection 
but also the terms that contain two fields with vanishing
VEVs (see Appendix C).

Before going into detail, it is worthwhile examining the NG modes
that are absorbed by the Higgs mechanism, because in some cases the 
vanishing eigenvalue in the mass matrices is not obvious.
There appear the following NG modes
\begin{enumerate}
\item ${\bf 16}+{\bf \overline{16}}+1$ of $SO(10)$ (namely,
$Q+U^c+E^c+h.c.+N^c$) 
in the breaking $E_6\rightarrow SO(10)$.
\item $Q+U^c+X+h.c.$ in the breaking 
$SO(10)\rightarrow SU(3)_C\times SU(2)_L\times SU(2)_R\times U(1)_{B-L}$.
\item $E^c+h.c.+N^c$ in the breaking
$SU(2)_R\times U(1)_{B-L}\rightarrow U(1)_Y$.
\end{enumerate}

First, we examine the mass matrix of ${\bf 24}$ in $SU(5)$.
Considering the additional term $A'^2$,
we write the mass matrices $M_I$, 
which correspond to the representations $I=G,W,X$:
\begin{equation}
M_I=\bordermatrix{
\bar I\backslash I  &    24_A    &     24_{A'}            \cr
24_A                &     0      & \alpha_I\lambda^{a'+a} \cr
24_{A'}   & \alpha_I\lambda^{a'+a}& \lambda^{2a'}         \cr
},
\end{equation}
where $\alpha_X=0$ and $\alpha_I\neq 0$ for $I=G,W$. 
One pair of $X$ is massless. This is absorbed by the Higgs mechanism.
The mass spectra are $(0, \lambda^{2a'})$ for $I=X$ and 
$(\lambda^{a'+a},\lambda^{a'+a})$ for $I=G,W$.

Next, we examine the mass matrices for the representation $I=Q, U^c$ 
and $E^c$, which are contained in ${\bf 10}$ of $SU(5)$. The mass 
matrices $M_I$ are written
\begin{equation}
\bordermatrix{
I\backslash \bar I  &\overline {16}_{\bar \Phi} & \overline {16}_{\bar C}&
                     \overline {16}_{A} &45_{A} &\overline {16}_{\bar C'}   & 
                    \overline {16}_{ A'} &  45_{A'}  \cr
16_\Phi & 0 & 0 & 0& 0& \lambda^{\bar c'+\phi}  & \lambda^{\phi+a'-\Delta\phi}
         & 0 \cr
16_C& 0 & 0 &0 & 0& \beta_I\lambda^{\bar c'+c} &  0  & 0 \cr
16_A & 0 & 0 & 0& 0& \lambda^{\bar c'+a+\Delta\phi}   & \lambda^{a'+a}
   & 0 \cr
45_{A} & 0 & 0 & 0 & 0& \lambda^{\bar c'+a+\Delta c}  & 0 & 
       \alpha_I\lambda^{a+a'}  \cr
16_{C'} & \lambda^{c'+\bar\phi} & \beta_I\lambda^{c'+\bar c} &
        \lambda^{a+c'-\Delta\phi} & \lambda^{a+c'-\Delta c} & 
        \lambda^{c'+\bar c'} & \lambda^{a'+c'-\Delta\phi} & 
        \lambda^{a'+c'-\Delta c}  \cr
16_{A'} & \lambda^{\bar \phi+a'+\Delta\phi} & 0 & 
        \lambda^{a+a'} & 0&\lambda^{\bar c'+a'+\Delta\phi} &  
        \lambda^{2a'}& \lambda^{2a'+\Delta\phi-\Delta c}  \cr
45_{A'} & 0 & 0 &0 & \alpha_I\lambda^{a'+a}
        & \lambda^{\bar c'+a'+\Delta c} &  
        \lambda^{2a'-\Delta\phi+\Delta c} 
        & \lambda^{2a'} \cr
},
\end{equation}
where we have used the relations
$\lambda^\phi\VEV{\Phi}\sim (\lambda^{\bar \phi}\VEV{\bar \Phi})^{-1}
\sim\lambda^{\Delta\phi}$ and
$\lambda^c\VEV{C}\sim(\lambda^{\bar c}\VEV{\bar C})^{-1}
\sim\lambda^{\Delta c}$ ($\Delta\phi=(\phi-\bar\phi)/2$, 
$\Delta c=(c-\bar c)/2$).
Since one pair of ${\bf \overline{16}}$ and ${\bf 16}$ (whose main 
modes are ${\bf \overline{16}}_{\bar \Phi}$ and ${\bf 16}_\Phi$) 
is absorbed by the Higgs mechanism in the process of breaking $E_6$ 
into $SO(10)$, we simply omit ${\bf 16}_\Phi$ and 
${\bf \overline{16}}_{\bar \Phi}$
 in deriving the mass spectrum. 
Then, the mass matrices can be written in the form of four 
$3\times 3$ matrices as
\begin{equation}
\left(\matrix{0 & A_I \cr B_I & C_I}\right),
\end{equation}
as in $SO(10)$ case.
It is obvious that the ranks of $A_I$ and $B_I$ reduce to two for 
$I=Q,U^c,E^c$ because
$(\alpha_I=0, \beta_I\neq 0)$ for $I=Q,U^c$ and
$(\alpha_I\neq 0, \beta_I=0)$ for $I=E^c$, where the vanishing of 
the parameter values is due to the fact that the NG modes are absorbed 
by the Higgs mechanism in the breaking
$SO(10)\rightarrow SU(3)_C\times SU(2)_L\times SU(2)_R\times
U(1)_{B-L}$ (for which the corresponding NG fields are $Q+U^c+h.c.$) 
and in the breaking $SU(2)_R\times U(1)_{B-L}$ into $U(1)_Y$
(for which the corresponding NG fields are $E^c+h.c.$).
The mass spectra become 
$(0,0,\lambda^{a'+a}, \lambda^{a'+a},\lambda^{c'+\bar c}, 
\lambda^{\bar c'+c}, \lambda^{2a'})$ for $I=Q,U^c$ and 
$(0,0,\lambda^{a'+a},\lambda^{a'+a},\lambda^{a'+a},
\lambda^{a'+a},\lambda^{\bar c'+c'})$ for $I=E^c$.

Finally, we examine the mass matrices of ${\bf 5}$ and ${\bf \bar 5}$
in $SU(5)$ and show how to realize the DT splitting.
Considering the additional terms,
we write the mass matrices $M_I$ for the representations 
$I=D^c(H_T),L(H_D)$ and their conjugates as
\begin{equation}
M_I=\left(\matrix{ 0 & A_I & 0 \cr  B_I & C_I & D_I \cr 
                                    E_I & F_I & G_I \cr}\right),
\end{equation}
\begin{equation}
A_I=\bordermatrix{
\bar I\backslash I   &  10_{C'} 
                    & 10_{\bar C'} &
                    \overline{16}_{\bar C'} &
                    \overline{16}_{A'}  \cr
10_\Phi & S_I\lambda^{c'+\phi+\Delta\phi} & \lambda^{\bar c'+\phi} & 
0  & 0 \cr
10_C  & 0 & \lambda^{\bar c'+c} & 0 & 0 \cr
16_C & 0 & 0 & \lambda^{\bar c'+c}
   & 0 \cr
16_A  & 0 & \lambda^{\bar c'+a+\Delta c}& \lambda^{\bar c'+a+\Delta\phi}
   & \lambda^{a'+a} \cr
}, 
\end{equation}
\begin{equation}
B_I=\bordermatrix{
\bar I\backslash I  & 10_\Phi&10_C &\overline{16}_{\bar C}&\overline{16}_{A}\cr
10_{C'} &S_I\lambda^{c'+\phi+\Delta\phi} & 0 & 0 &\lambda^{c'+a-\Delta c} \cr
10_{\bar C'} &\lambda^{\bar c'+\phi} &\lambda^{\bar c'+c} &
\lambda^{\bar c'+\bar c-\Delta c} & \lambda^{\bar c'+a-\Delta\phi-\Delta c} \cr
16_{C'} & 0 & 0 &\lambda^{c'+\bar c} &\lambda^{a+c'-\Delta\phi} \cr
16_{A'}& 0 & 0 & 0 &\lambda^{a'+a} \cr
}, 
\end{equation}
\begin{equation}
C_I=\bordermatrix{
\bar I\backslash I  & 10_{C'} 
                    & 10_{\bar C'}  &
                    \overline{16}_{\bar C'} &
                    \overline{16}_{A'}  \cr
10_{C'}  & \lambda^{2c'+\Delta\phi} & 
  \lambda^{\bar c'+c'} &
  \lambda^{\bar c'+c'+\Delta\phi-\Delta c}  & \lambda^{c'+a'-\Delta c} \cr
10_{\bar C'}  &\lambda^{\bar c'+c'} & 
\lambda^{2\bar c'-\Delta\phi} & \lambda^{2\bar c'-\Delta c} & 
\lambda^{\bar c'+a'-\Delta\phi-\Delta c} \cr
16_{C'} & \lambda^{2c'+\Delta c} & 
  \lambda^{\bar c'+c'-\Delta\phi+\Delta c} &\lambda^{\bar c'+c'} &  
\lambda^{a'+c'-\Delta\phi} \cr
16_{A'}& \lambda^{c'+a'+\Delta\phi+\Delta c} & 
\lambda^{\bar c'+a'+\Delta c} & \lambda^{\bar c'+a'+\Delta\phi} &  
\lambda^{2a'} \cr
},
\end{equation}
\begin{equation}
D_I=\bordermatrix{
\bar I\backslash I  & 10_{\bar \Phi} &10_{\bar C} &\overline{16}_{\bar \Phi} \cr
10_{C'} &\lambda^{c'+\bar\phi} &\lambda^{c'+\bar c} & 
   \lambda^{c'+\bar\phi+\Delta\phi-\Delta c} \cr
10_{\bar C'} &\lambda^{\bar c'+\bar\phi-\Delta\phi} & 
       \lambda^{\bar c'+\bar c-\Delta\phi} & 
       \lambda^{\bar c'+\bar\phi-\Delta c} \cr
16_{C'} & \lambda^{c'+\bar\phi-\Delta\phi+\Delta c} &
      \lambda^{c'+\bar c-\Delta\phi+\Delta c} & \lambda^{c'+\bar\phi} \cr
16_{A'} & 0 & 0 & \lambda^{\bar\phi+a'+\Delta\phi} \cr
}, 
\end{equation}
\begin{equation}
E_I=
\bordermatrix{
\bar I\backslash I  &10_\Phi & 10_C& 
                    \overline{16}_{\bar C} &\overline{16}_{A}   \cr
10_{\bar \Phi} & 
0 & 0& 0 & 
\lambda^{\bar\phi+a-\Delta\phi-\Delta c}   \cr
10_{\bar C} &
0 & 0& 0 & 0
 \cr
16_\Phi & 0 & 0 &  0 & 0  \cr
},
\end{equation}
\begin{equation}
F_I=\bordermatrix{
\bar I\backslash I   &  10_{C'} 
                    & 10_{\bar C'} &
                    \overline{16}_{\bar C'} &
                    \overline{16}_{A'}  \cr
10_{\bar \Phi}  &
\lambda^{c'+\bar\phi} & \lambda^{\bar c'+\bar\phi-\Delta\phi}&
\lambda^{\bar c'+\bar\phi-\Delta c} & 
 \lambda^{\bar\phi+a'-\Delta\phi-\Delta c} \cr
10_{\bar C}  & \lambda^{\bar c+c'} &
\lambda^{\bar c'+\bar c-\Delta\phi}& \lambda^{\bar c'+\bar c-\Delta c} & 
\lambda^{\bar c+a'-\Delta\phi-\Delta c} \cr
16_\Phi  & 0 & \lambda^{\bar c'+\phi-\Delta\phi+\Delta c}& 
   \lambda^{\bar c'+\phi}
   & \lambda^{a'+\phi-\Delta\phi} \cr
}, 
\end{equation}
\begin{equation}
G_I=\bordermatrix{
\bar I\backslash I   &  10_{\bar \Phi} 
                    & 10_{\bar C} &
                    \overline{16}_{\bar \Phi} \cr
10_{\bar \Phi}  &
\lambda^{2\bar\phi-\Delta\phi} & \lambda^{\bar \phi+\bar c-\Delta\phi}&
\lambda^{2\bar\phi-\Delta c} \cr
10_{\bar C}  & \lambda^{\bar \phi+\bar c-\Delta\phi} & 0 & 0 \cr
16_\Phi  & 0 & 0 & 0 \cr
}, 
\end{equation}
where $S_{D^c}\neq 0$ and $S_L=0$.
It is obvious that the rank of $A_L$ is three, which is smaller than
the rank of $A_{D^c}$. This implies that the rank of $M_L$ is smaller
than the rank of $M_{D^c}$, and actually 
the rank of the matrix $M_I$ is 10 for $I=D^c$
and 9 for $I=L$. One pair of massless fields, 
${\bf 16}$ and ${\bf \overline{16}}$ (whose main modes are 
${\bf 16}_{\Phi}$ and ${\bf \overline{16}}_{\bar \Phi}$), 
is the NG mode, which is absorbed by the Higgs mechanism in the 
breaking of $E_6$ into $SO(10)$. 
The other massless mode for $I=L$ is the so-called doublet
Higgs. The massless mode is given by
\begin{eqnarray}
H_u&\sim&\bar L({\bf 10}_\Phi)+\lambda^{\phi-c}\bar L({\bf 10}_C), \\
H_d&\sim&L({\bf 10}_\Phi)+\lambda^{\phi-c}L({\bf 10}_C).
\end{eqnarray}

As noted above, ${\bf 16}_{\Phi}$ and ${\bf \overline{16}}_{\bar \Phi}$
are absorbed by the Higgs mechanism, and ${\bf 10}_{\bar \Phi}$ and
${\bf 10}_{\bar C}$ can become massive through the matrix $G_I$, 
whose elements are generally larger than the elements of $D_I$, 
$E_I$ and $F_I$.
Their masses become
$(\lambda^{\bar \phi+\bar c-\Delta\phi},
\lambda^{\bar \phi+\bar c-\Delta\phi})$.
We simply ignore the matrices $D_I$, $E_I$, $F_I$ and $G_I$
in the following argument.
Since the elements of $A_I$ and $B_I$ are generally larger than those
of $C_I$, we can estimate the mass spectrum of the other modes of
$D^c$ from $A_{D^c}$ and $B_{D^c}$
as $(\lambda^{c'+\phi+\Delta\phi},\lambda^{c'+\phi+\Delta\phi},
\lambda^{\bar c'+c},\lambda^{\bar c'+c},\lambda^{ \bar c'+ c},
\lambda^{ c'+\bar c},\lambda^{a'+a},\lambda^{a'+a})$, and
the mass spectrum of the other modes of $L$ as
$(0,\lambda^{\bar c'+c},\lambda^{\bar c'+c},\lambda^{ \bar c'+ c},
\lambda^{ c'+\bar c},\lambda^{a'+a},\lambda^{a'+a},
\lambda^{2c'+\Delta\Phi})$.
It is obvious that to realize proton decay, we have to pick up an  
element of $C_I$. Since such an  element is generally smaller 
than the mass scale of $D^c$, proton decay is suppressed.
The effective colored Higgs mass is estimated as
$(\lambda^{c'+\phi+\Delta\phi})^2/\lambda^{2c'+\Delta\phi}=
\lambda^{2\phi+\Delta\phi}$, 
which is usually larger than the cutoff scale.
For example, for the typical charge assignment in Table II, 
$2\phi+\Delta\phi=-17/2$.

According to the above argument, the mass spectrum of the 
superheavy particles are determined only by the anomalous 
$U(1)_A$ charges, so we can examine whether coupling unification 
is realized or not. 
Before going into discussion of this point (given in the next 
subsection), we define the reduced 
mass matrices $\bar M_I$ by getting
rid of the massless modes from the original mass matrices $M_I$.
The ranks of the reduced matrices in our $E_6$ model are
$\bar r_{X}=1$, $\bar r_G=\bar r_W=2$, 
$\bar r_{Q}=\bar r_{U^c}=\bar r_{E^c}=5$, $\bar r_L=9$ and 
$\bar r_{D^c}=10$.
It is interesting that the determinants of the reduced mass matrices are
evaluated mainly as simple sums of the anomalous $U(1)_A$ charges of
massive modes:
\begin{eqnarray}
\det \bar M_I (I=G,W)&=& \lambda^{2(a+a')} \\
\det \bar M_X&=&\lambda^{2a'} \\
\det \bar M_I(I=Q,U^c)&=& \lambda^{2a+4a'+c+\bar c+c'+\bar c'}\\
\det \bar M_{E^c}&=&\lambda^{4a+4a'+c'+\bar c'} \\
\det \bar M_D^c&=&\lambda^{3(c+\bar c+c'+\bar c')+2(a+a'+\phi+\bar \phi)} \\
\det \bar M_L&=&\lambda^{3(c+\bar c+c'+\bar c')+2(a+a'+\bar \phi)-\Delta\phi}.
\end{eqnarray}
Note that the last equation for $\det \bar M_L$ is not determined 
by a simple sum of the charges of massive modes.
 The difference is $-\Delta\phi$. This is because the masses of the 
${\bf 10}$ representation of $SO(10)$ in the multiplets 
$X_i({\bf 27})$ of $E_6$ are derived as 
$\lambda^{x_i+x_j+\Delta\phi}$ from the terms $X_iX_j\Phi$ with VEV
$\VEV{\Phi}\sim\lambda^{-\frac{1}{2}(\phi+\bar \phi)}$, which is 
different from the naive expectation, $\lambda^{x_i+x_j}$.
In order to calculate the elements of mass matrices, it is useful 
to introduce the following `effective' charges:
\begin{eqnarray}
&&x_i(10(5+\bar 5),27)\equiv x_i+\frac{1}{2}\Delta\phi,\quad
\bar x_i(10(5+\bar 5),\overline{27})\equiv \bar x_i-\frac{1}{2}\Delta\phi, \\
&&x_i(16(\bar 5),27)\equiv x_i+\Delta c-\frac{1}{2}\Delta\phi,\quad
\bar x_i(\overline{16}(5),\overline{27})\equiv 
\bar x_i-\Delta c+\frac{1}{2}\Delta\phi, \\
&&x_i(16(10),27)\equiv x_i, \quad
\bar x_i(\overline{16}(\overline{10}),\overline{27})\equiv \bar x_i, \\
&&a(16(\bar 5),78)\equiv a+\Delta c+\frac{1}{2}\Delta\phi,\quad
a(\overline{16}(5),78)\equiv a-\Delta c-\frac{1}{2}\Delta\phi, \\
&&a(16(10),78)\equiv a+\Delta\phi,\quad
a(\overline{16}(\overline{10}),78)\equiv a-\Delta\phi,\\
&&a(45(10),78)\equiv a+\Delta c,\quad
a(45(\overline{10}),78)\equiv a-\Delta c, \\
&&a(24,78)\equiv a.
\end{eqnarray}
Here, the effective charges $x_i(10(5+\bar 5),27)$
are for ${\bf 10}$ of $SO(10)$ from $X_i({\bf 27})$ of $E_6$ and
$\bar x_i(10(5+\bar 5),\overline{27})$ are 
for ${\bf 10}$ of $SO(10)$ from $\bar X_i({\bf \overline{27}})$ of $E_6$,
etc.

We thus find that all the elements of the mass matrices can be 
computed as simple sums of the effective charges 
of superheavy particles if they are not vanishing, 
and the determinants of the mass matrices are
also determined by simple sums of the effective charges.
We will use this result in calculating the running gauge couplings
below.

\subsection{Coupling unification}
In this subsection, we apply the general analysis of gauge coupling 
unification given in Ref.~\cite{maekawa3} to our scenario.

The pattern of the $E_6$ breaking in our model is as follows.
At the scale $\Lambda_\Phi\sim \lambda^{-(\phi+\bar \phi)/2}$, 
$E_6$ is broken into $SO(10)$. $SO(10)$ is broken into
$SU(3)_C\times SU(2)_L\times SU(2)_R\times U(1)_{B-L}$ at the scale
$\Lambda_A\sim \lambda^{-a}$, which is broken into the standard gauge
group at the scale $\Lambda_C\sim \lambda^{-(c+\bar c)/2}$.

In this paper, we carry out analysis based on the renormalization group
equations up to one loop.%
\footnote{Since we ignore the order 1 coefficients,
a higher-order calculation does not improve the accuracy.}
The conditions of the gauge coupling unification are given by
\begin{equation}
\alpha_3(\Lambda_A)=\alpha_2(\Lambda_A)=
\frac{5}{3}\alpha_Y(\Lambda_A)\equiv\alpha_1(\Lambda_A),
\end{equation}
where 
$\alpha_1^{-1}(\mu>\Lambda_C)\equiv 
\frac{3}{5}\alpha_R^{-1}(\mu>\Lambda_C)
+\frac{2}{5}\alpha_{B-L}^{-1}(\mu>\Lambda_C)$.
Here $\alpha_X=\frac{g_X^2}{4\pi}$ and 
the parameters $g_X (X=3,2,R,B-L,Y)$ are the gauge couplings of 
$SU(3)_C$, $SU(2)_L$, $SU(2)_R$, $U(1)_{B-L}$ and $U(1)_Y$, 
respectively.

Using the fact that the three gauge couplings of MSSM meet at the 
scale $\Lambda_G\sim 2\times 10^{16}$ GeV,
the above conditions for gauge coupling unification can be rewritten
\begin{eqnarray}
&&b_1\ln \left(\frac{\Lambda_A}{\Lambda_G}\right)
+\sum_I\Delta b_{1I}\ln \left(\frac{\Lambda_A^{\bar r_I}}{\det \bar M_I}
\right)
-\frac{12}{5}\ln \left(\frac{\Lambda_A}{\Lambda_C}\right) \label{alpha1}
\nn\\
&=&b_2\ln \left(\frac{\Lambda_A}{\Lambda_G}\right)
+\sum_I\Delta b_{2I}\ln\left(\frac{\Lambda_A^{\bar r_I}}{\det \bar M_I}
\right) \\
&=&b_3\ln \left(\frac{\Lambda_A}{\Lambda_G}\right)
+\sum_I\Delta b_{3I}\ln\left(\frac{\Lambda_A^{\bar r_I}}{\det \bar M_I}
\right),
\end{eqnarray}
where $(b_1,b_2,b_3)=(33/5,1,-3)$ are the 
renormalization group coefficients
for MSSM
and $\Delta b_{aI}(a=1,2,3)$ are the corrections to the 
coefficients from the massive fields 
$I=Q+\bar Q,U^c+\bar U^c, E^c+\bar E^c, D^c+\bar D^c,
L+\bar L, G, W$ and $X+\bar X$.
The last term in Eq. (\ref{alpha1}) is from the breaking 
$SU(2)_R\times U(1)_{B-L}\rightarrow U(1)_Y$ by the VEV $\VEV{C}$.
Since all the mass matrices and the symmetry breaking scale appearing 
in the above conditions are determined by the anomalous $U(1)_A$ 
charges, these conditions can be translated into a constraint 
on the effective charge of the doublet Higgs fields,
\begin{equation}
\phi+\frac{1}{2}\Delta\phi\sim 0,
\label{h}
\end{equation}
and into a condition on the cutoff scale,
\begin{equation}
\Lambda\sim \Lambda_G.
\end{equation}
As discussed in Ref. \cite{maekawa3}, this a is quite general result 
and independent of the details of the Higgs sector. The essential 
point is that only the charges of massless modes are important to 
determine whether coupling unification is realized, and all 
other effects are cancelled out in the unification conditions, 
except the charge of the doublet Higgs.
Note that the condition (\ref{h}) does not require
$\phi+\frac{1}{2}\Delta\phi= 0$. Actually, even with a typical charge
assignment, in which $\phi+\frac{1}{2}\Delta\phi= -4.25$, the coupling
unification is realized, using the ambiguities of the order 1 
coefficients(see Fig. 1).

Since the unification scale
$\Lambda_U\sim \lambda\Lambda_G$ is smaller than the usual GUT scale
$\Lambda_G\sim 2\times 10^{16}$ GeV, proton decay via dimension 6 
operators $p\rightarrow e^+\pi^0$ may be seen in the near future. 
If we roughly estimate the lifetime of the proton using the formula 
in Ref.~\cite{hisano} and a recent result from a lattice calculation 
for the hadron matrix element parameter
$\alpha$~\cite{lattice}, the lifetime of the proton in our scenario
becomes
\begin{equation}
\tau_p(p\rightarrow e^+\pi^0)\sim 2.8\times 10^{33}\left(\frac{\Lambda_U}
{5\times 10^{15}~{\rm GeV}}\right)^4
\left(\frac{0.015~{\rm GeV}^3}{\alpha}\right)^2  {\rm years},
\end{equation}
because the unification scale is around $5\times 10^{15}$ GeV.
It is interesting that this value of the lifetime is just around the 
present experimental lower bound
\cite{SKproton}
\begin{equation}
\tau_{exp}(p\rightarrow e^+\pi^0)>2.9\times 10^{33}~{\rm years}.
\end{equation}
Of course, since we have ambiguities in the order 1 coefficients and 
the hadron matrix element parameter $\alpha$, and because 
the lifetime of the proton is strongly dependent on the GUT scale and 
this parameter, this prediction may not be very reliable. 
However, this rough estimation provide strong motivation for
experiments to detect proton decay, 
because the lifetime of a nucleon via dimension 6 operators
must be less than that in the usual SUSY GUT scenario. 

We should comment on proton decay via dimension 5 operators.
The effective colored Higgs mass is given by
$\lambda^{2\phi+\Delta\phi}\Lambda$, so the experimental 
constraint requires $2\phi+\Delta\phi\leq -3$. 
With 
the typical charge assignment in Table II, the effective colored Higgs
mass is around $\sim 10^{22}$ GeV, so that proton decay via dimension 
5 operators is suppressed. 

\subsection{How to determine charges}
It is worthwhile explaining the method for determining 
the symmetry and quantum numbers in the Higgs sector to realize 
DT splitting. 
There are several terms which must be forbidden in order to realize 
DT splitting:
\begin{enumerate}
\item $\Phi^3$, $\Phi^2C$, $\Phi^2C'$, $\Phi^2C'Z$ induce a large mass 
of the doublet Higgs.
\item $\bar CA'C$,$\bar CA'AC$,$\bar \Phi A'\Phi$ would destabilize the
DW form of $\VEV{A}$. 
\item $\bar \Phi A'C$, $\bar CA'\Phi$, $\bar \Phi A'AC$, $\bar CA'A\Phi$, 
$\bar \Phi A'ZC$, $\bar C A'Z\Phi$ lead to the undesired VEV 
$\VEV{{\bf 16}_C}=0$, 
unless  another singlet field is introduced. 
\item $A'A^n(n\geq 4)$ make it less natural to obtain a DW-type of VEV.
\end{enumerate}
Most of these terms can be easily forbidden
by the SUSY zero mechanism. For example, if we choose $\phi<0$, 
then $\Phi^3$ is forbidden, and if we choose $\bar c+c+a+a'<0$, 
then $\bar CA'AC$ is forbidden.
Once these terms are forbidden by the SUSY zero mechanism, 
higher-dimensional terms, which also become dangerous (for example, 
$\bar CA'A^3C$ and $\bar CA'C\bar CAC$), are automatically forbidden, 
as in $SO(10)$ cases.
Contrastingly, the following terms are necessary:
\begin{enumerate}
\item $A'A$, $\bar \Phi A'A^3\Phi$ to obtain a DW-type VEV $\VEV{A}$.
\item $\Phi^2AC'$ for doublet-triplet splitting.
\item $\bar C'(A+Z)C$, $\bar C(A+Z)C'$ for alignment between 
the VEVs $\VEV{A}$ and $\VEV{C}$ and to give superheavy masses 
to the PNGs.
\item $\bar\Phi A'A\Phi$ for alignment between 
the VEVs $\VEV{A}$ and $\VEV{\Phi}$ and to give the superheavy masses 
to the PNGs.
\item $\bar \Phi^3$,$\bar \Phi^2\bar C$ to give superheavy masses to 
two ${\bf 10}$ of $SO(10)$, which save the number of fields with 
positive charges.
\end{enumerate}

In order to forbid $\Phi^2C'$ but not $\Phi^2AC'$, 
we have to introduce $Z_2$ parity. The same $Z_2$ parity can forbid
$\bar\Phi A'\Phi$, while allowing the term $\bar\Phi A'A\Phi$. 
We have some ambiguities to assign the $Z_2$ parity, but once the 
parity is fixed, the above requirements simply become inequalities.
In addition to these inequalities, we require that the total charges 
of the operators $A$, $\bar \Phi\Phi$ and $\bar CC$ be negative. 
If, as discussed in the previous section, we adopt $a=-1$ to realize 
proton stability, then the inequalities $a'+3a+\phi+\bar \phi\geq 0$
and $a'+5a<0$ lead to $\phi+\bar \phi=-1$ and $a'=4$. The relation
$\bar\phi+\phi=-1$ means that the gauge singlet
operator $\bar\Phi\Phi$ can be regarded as the FN field $\Theta$, 
as discussed in \S2.
The other inequalities are easily satisfied.

Of course, the above stated conditions are necessary but not 
sufficient.  As in the previous subsection, we have to write down 
the mass matrices of the Higgs sector to know whether an assignment 
actually works or not.

\section{Constraints from the matter sector}
For the matter fields, we introduce three ${\bf 27}$, 
$\Psi_i$ ($i=1,2,3$).
As discussed in Refs. \cite{maekawa,BM}, we adopt 
$(\psi_1,\psi_2,\psi_3)=(3+n,2+n,n)$, and the charge of the Higgs 
$\phi=-2n$ in order to realize a CKM matrix and a large top Yukawa 
coupling of $O(1)$.
As discussed in the previous section, the cutoff scale $\Lambda$ 
must be around the usual GUT scale 
$\Lambda_G\sim 2\times 10^{16}$ GeV. 
This requires $a\leq -1$. If we have an integer charge for $a$, 
then we must adopt $a=-1$. 
As discussed in Ref.~\cite{BM}, the conditions for obtaining realistic
quark and lepton mass matrices with bi-large neutrino mixing angles are
\begin{equation}
c-\bar c=\phi-\bar\phi+1=2n-9-l,
\end{equation}
where there is some ambiguity in the parameter $l$, but as
discussed in Ref.~\cite{BM}, for models with $\Lambda\sim \Lambda_G$,
we adopt $l=-1$ or $-2$.%
\footnote{
Using the ambiguities of the order 1 coefficients, a rather larger 
range of values of the parameter $l$ may be allowed, for example, 
$l=-3,-4$. But in the following discussion, these larger ambiguities 
do not change the result.
}
Since the condition for gauge coupling unification is 
$\phi+\frac{1}{4}(\phi-\bar \phi)=\frac{1}{4}(-6n-10-l)\sim 0$,
a small value of $n$ is required. From the condition 
$\phi+\bar \phi=-6n+10+l\leq  -1$, the smallest value of 
$n$ becomes $3/2$ for $l=-2$. 
Then, we obtain $\phi=-3$, $\bar \phi=2$ and
$c-\bar c=-4$. It is non-trivial that the conditions for 
bi-large neutrino mixing angles and for gauge coupling unification 
lead to $\bar \phi+\phi=-1$, which is required for DT splitting.
Since $\phi+\bar\phi=-1$, we must adopt $a'=4$ to allow
the important term $\bar \Phi A'A^3\Phi$ and to forbid the term 
$A'A^5$. 
Then, in order to forbid the terms
$\bar \Phi A'AC$, $\bar CA'A\Phi$ and $\bar CA'AC$, we have to take
$c<-5$ and $\bar c<0$. On the other hand, since the term 
$\Psi_1\Psi_3C$ is required to realize realistic mass matrices of 
the quark and lepton, as discussed in Ref.~\cite{BM}, $c\geq -6$ is 
needed. Hence we must take $c=-6$ and
$\bar c=-2$. It is interesting that the economical condition for the 
$\mu$ problem~\cite{maekawa2,maekawa3}, 
\begin{equation}
-1\leq 2\phi-(c+\bar c)+\frac{1}{2}(\phi-\bar \phi)\leq 1,
\label{mu}
\end{equation}
is automatically satisfied and the required terms $\bar\Phi^3$ and 
$\bar\Phi^2\bar C$ happen to be allowed.
We have some freedom in choosing the charges $z$, $c'$ and $\bar c'$. 
If we take $z=-2$, then we must adopt $c'=7$, because the term 
$ZC'\Phi^2$ must be forbidden, while the term $C'A\Phi^2$ must be 
allowed. 
Also, $\bar c'\geq 8$ is required to obtain the term $\bar C'(A+Z)C$. 
In the typical charge assignment, we adopt the minimal value 
$\bar c'=8$.%
\footnote{
In order to solve the $\mu$ problem economically, we need the term 
$\bar C'\Phi Z(\Phi^3)$ or $\bar C'\Phi (\Phi^3)$ when we adopt 
even $Z_2$ parity for $C'$. This leads to $\bar c'\geq 14$ or
$\bar c'\geq 12$. 
However, in these cases, the gauge couplings seem to diverge below
the GUT scale. Therefore, the additional gauge singlet field $S$ 
with positive charge $s\geq 3\phi$ may be required.
}

The charges of the matter sector 
$\Psi_i({\bf 27})~(i=1,2,3)$ become half integers as
$(\psi_1,\psi_2,\psi_3)=(9/2,7/2,3/2)$ in this case.
It is interesting that we do not have to introduce 
R-parity, because
half integer anomalous $U(1)_A$ charges can play the same role.

\begin{figure}[htb]
\begin{center}
\leavevmode
\epsfxsize=110mm
\put(300,50){{\large $\bf{\log \mu (GeV)}$}}
\put(0,260){{\Large $\bf{\alpha^{-1}}$}}
\put(29,240){$\alpha_1^{-1}$}
\put(31,150){$\alpha_2^{-1}$}
\put(31,90){$\alpha_3^{-1}$}
\epsfbox{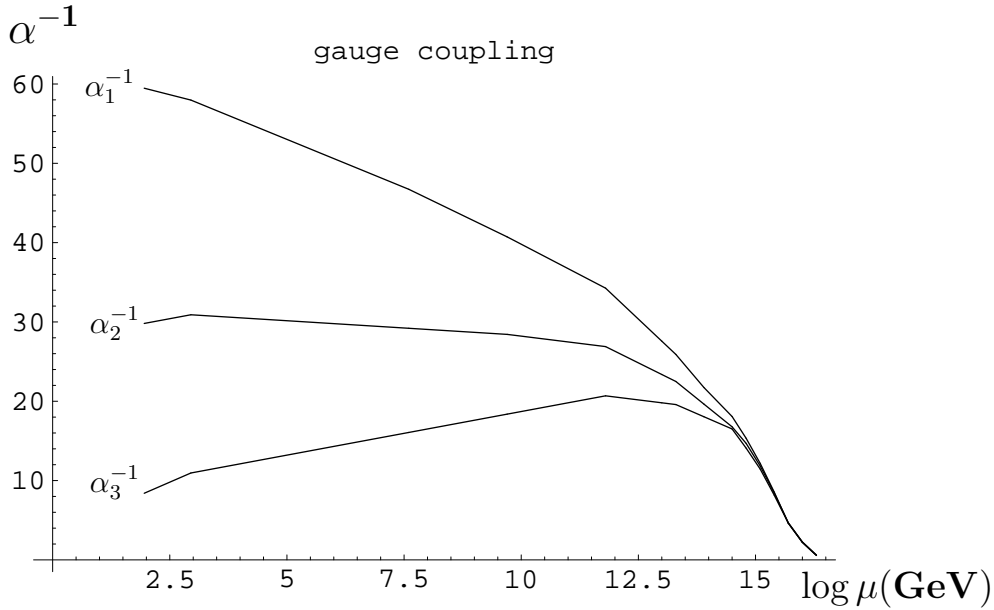}
\vspace{-2cm}
\caption{
Here we adopt $\lambda=0.25$, 
$\alpha_1^{-1}(M_Z)=59.47$, $\alpha_2^{-1}(M_Z)=29.81$,
$\alpha_3^{-1}(M_Z)=8.40$, and the SUSY breaking scale 
$m_{SB}\sim 1$ TeV.
We also use the anomalous $U(1)_A$ charges $a'=4$, $a=-1$, 
$(\psi_1,\psi_2,\psi_3)=(9/2,7/2,3/2)$, $\phi=-3$, $\bar \phi=2$, 
$c=-6$, $\bar c=-2$, $z=-2$, $c'=7$ and $\bar c'=8$.
Using the ambiguities of coefficients expressed by $0.5\leq y\leq 2$,
the three gauge couplings meet at around 
$\lambda^{-a}\Lambda_G\sim 5\times 10^{15}$ GeV. 
}
\label{fig_1}
\end{center}
\end{figure}

From the above consideration, all the charges are fixed except the 
singlets. Terefore we can calculate the running flows of the gauge 
couplings (see Fig. \ref{fig_1}). 
Here we use the ambiguities of the coefficients expressed by 
$0.5\leq y\leq 2$.
It is shown that the three gauge couplings actually meet around
$\lambda^{-a}\Lambda_G\sim 5\times 10^{15}$ GeV. Note that
the unified gauge coupling at the cutoff scale is still finite, 
although it is so large that the perturbative calculation is not 
reliable.
Since the value of the unified gauge coupling is strongly dependent 
on the actual charge assignment and the value $\lambda$ (which is 
weakly dependent on the coefficients of order 1), we expect that 
it is large but finite at the cutoff scale. 

\section{Discussion and summary}
We emphase that the effective $SO(10)$ theory that 
is obtained with the non-vanishing VEVs $\VEV{\Phi}=\VEV{\bar\Phi}$ 
from $E_6$ GUT is generally different from the $SO(10)$ GUT 
with anomalous $U(1)_A$ gauge symmetry.
This is because the VEV 
$\VEV{\Phi}\sim \lambda^{-\frac{1}{2}(\phi+\bar\phi)}$ 
is generally different from the naively expected value, 
$\VEV{\Phi}\sim \lambda^{-\phi}$. For example, the mass of 
${\bf 10}_F$ in $SO(10)$ is obtained from the term 
$\lambda^{2f+\phi}F({\bf 27})^2\Phi({\bf 27})$ by developing the
VEV $\VEV{\Phi}\sim \lambda^{-\frac{1}{2}(\phi+\bar \phi)}$ as
$m_F\sim \lambda^{2f+\frac{1}{2}(\phi-\bar\phi)}$, which is different
from the naively expected value, $m_F\sim \lambda^{2f}$.
This is the same effect as that discussed in Ref.~\cite{maekawa} with 
regard to the right-handed neutrino mass in $SO(10)$ GUT.  
The right-handed neutrino mass has been estimated as 
$m_{\nu_F}\sim \lambda^{2f+\frac{1}{2}(c-\bar c)}$,
which is different from the naively expected value, $\lambda^{2f}$. 
Therefore, it is obvious that the $SO(10)$ or $E_6$ GUT with 
anomalous $U(1)_A$ gauge symmetry
is essentially different from the MSSM with anomalous $U(1)_A$ gauge
symmetry. 
Note that the difference is caused by $\Delta\phi$ and $\Delta c$.

In this paper, we have proposed a realistic Higgs sector in 
$E_6$ grand unified theory, which can realize DT splitting and proton
stability. In this scenario, the anomalous $U(1)_A$ gauge symmetry 
plays a critical role.
Moreover, using the matter sector in $E_6$ GUT given 
in Ref. \cite{BM},
we have proposed a completely consistent $E_6$ GUT scenario.
Since we have introduced all the interactions that are allowed by
the symmetry, the model can be defined by the symmetry and the 
quantum numbers of the fields. In our scenario, after deciding
the field content (Higgs and matter), the model can be defined with 
only 8 integer charges (+3 charges for singlet fields)
in the Higgs sector and 3 (half-)~integer charges in the matter 
sector.
It is quite intriguing that by choosing only 11 (half-) integer 
charges (+3 charges for singlet fields),
not only are DT splitting with proton stability and gauge coupling
unification realized but also the realistic structure of 
quark and lepton mass matrices, 
including bi-large neutrino mixing, are obtained.
Moreover, the FCNC process is automatically suppressed. 
In other words, the charge assignment, which is almost determined 
by realizing realistic
quark and lepton mass matrices and gauge coupling unification, 
also solves the DT splitting with proton stability. 
For example, the relation $\bar \phi+\phi=-1$ is required to obtain 
realistic quark 
and lepton mass matrices and gauge coupling unification, and it is 
also independently needed to realize DT splitting.
Moreover, since the half-integer charges of the matter sector 
play the same role as $R$-parity, we do not have to introduce 
$R$-parity.

Of course if it is allowed for the half-integer charges to be 
assigned to the Higgs sector, there are other possibilities.
For example, we can adopt another charge assignment, 
in which the half integer charges
play the same role as $Z_2$ parity. If we use
$a=-1/2,a'=5/2,\phi=-3,\bar\phi=2,c=-5,\bar c=-1,c'=13/2,\bar c'=13/2,
z_i=-i/2(i=3,7,11),
\psi_1=9/2,\psi_2=7/2,\psi_3=3/2$ with odd $R$-parity in the matter 
sector $\Psi_i$, we can obtain the completely consistent 
$E_6$ GUT again.
Since the absolute values of the charges of this model are smaller 
than those of the previous model, the unified gauge coupling 
at the unified scale $\lambda^{-a}$ becomes smaller than that of 
the previous model.
Therefore, the unified gauge coupling at the cutoff scale must be
finite in this model.
Then, since the unification scale $\lambda^{-a}$ is larger than 
that of previous model, the model predicts a longer lifetime of 
the nucleon, which is roughly estimated as
\begin{equation}
\tau_p(p\rightarrow e^+\pi^0)\sim 4.5\times 10^{34}\left(\frac{\Lambda_U}
{10^{16}~{\rm GeV}}\right)^4
\left(\frac{0.015~{\rm GeV}^3}{\alpha}\right)^2  {\rm years}.
\end{equation}
Though this predicted value is significantly longer than the present 
experimental lower bound, we hope that the next generation of 
experiments can reach this value.%
\footnote{
Unfortunately, since the term $A'A^5$ is allowed by the symmetry 
in this charge assignment,
it is less natural to obtain a DW-type VEV in this model than in 
the previous model.
However, the number of VEVs is still finite.
}

Though the requirement on $E_6$ GUT is so severe that the possible 
charge assignments are fairly restricted, there are several 
possibilities for this assignment.
However, since our scenario requires only several integer charges
as the input paramters to provide realistic results both in 
the matter sector and in the Higgs sector, 
we believe that this scenario indeed describs our world.

\section{Acknowledgement}
We would like to thank M. Bando and T. Kugo for their collaboration 
in the early stages of this work. We also thank M. Bando for reading 
the manuscript and making valuable comments that made this manuscript 
more easily understandable.


\appendix

\section{Factorization}
  As mentioned in \S4, the naive extension
of DT splitting in $SO(10)$ GUT into $E_6$ GUT
does not work.
  In the $SO(10)$ DT splitting, the interaction
$(A^\prime A)_{\bf 54}(A^2)_{\bf 54}$ plays an essential role.
  In $E_6$ GUT, however, the term $A^\prime A^3$
does not include the interaction 
$({\bf 45}_{A'}{\bf 45}_A)_{\bf 54}({\bf 45}_A^2)_{\bf 54}$.
  Therefore the superpotential
\begin{equation}
W_{A^\prime} =  \lambda^{a^\prime+a}\alpha A^\prime A
              + \lambda^{a^\prime+3a} ( \beta(A^\prime A)_{\bf 1}
                                             (A^2)_{\bf 1}
                                      + \gamma(A^\prime A)_{\bf 650}
                                              (A^2)_{\bf 650})
\end{equation}
does not realize the DW VEV naturally.
  Here, we show that the term $A^\prime A^3$ of $E_6$ actually does not
include the interaction $({\bf 45}_{A'}{\bf 45}_A)_{\bf 54}
({\bf 45}_A^2)_{\bf 54}$ of $SO(10)$.

  The VEV of $SO(10)$ adjoint Higgs can be represented in the form
$\VEV{A}=\tau_2\times {\rm diag}(x_1,x_2,x_3,x_4,x_5)$, because of the
$SO(10)$ rotation and D-flatness condition (see Appendix
\ref{D-flatness}).
  In this gauge,
\beqn
  A^\prime A &=& 2 \sum_i x^\prime_i x_i ,\\
  (A^\prime A)_{\bf 54}(A^2)_{\bf 54} &=& 2 \sum_i x^\prime_i {x_i}^3 
  -\frac{2}{5}\left(\sum_i x'_ix_i\right)\left(\sum_jx_j^2\right).
\eeqn
  In the same manner, the VEV of $E_6$ adjoint Higgs can be
represented in the form $\VEV{{\bf 1}_A}=y$ , $\VEV{{\bf 16}_A}=
\VEV{{\bf \overline{16}}_A}=0$, $\VEV{{\bf 45}_A}=\tau_2\times
{\rm diag}(x_1,x_2,x_3,x_4,x_5)$.
  In this gauge, the VEV $\VEV A$ can be represented
as $27 \times 27$ matrix as 
\begin{equation}
  \VEV{A} = \left(
  \begin{array}{ccc}
    {2\over \sqrt3}y & 0 & 0 \\
    0 & \theta^{MN}{T_{16}}^{MN}
        + {1\over2\sqrt3}y{\bf 1}_{16} & 0 \\
    0 & 0 & \theta^{MN}{T_{10}}^{MN}
            - {1\over\sqrt3}y{\bf 1}_{10} \\
\end{array}
\right)\label{A-matrix}.
\end{equation}
  Here, ${T_{i}}^{MN}$ is the $i \times i$ matrix representation of
$SO(10)$ generators and the summation of the indices $M$ and $ N$ is 
understood from 1 to 10 with  $M>N$. 
  Also, ${\bf 1}_{i}$ is the $i \times i$ unit matrix. 
  Explicitly, we have 
\beqn
  ({T_{10}}^{MN})_{KL} &=& -i( \delta^M_K\delta^N_L
                             - \delta^M_L\delta^N_K ), \\
  ({T_{16}}^{MN})_{\alpha\beta}
                       &=& {1\over2}(\sigma^{MN})_{\alpha\beta} \nn\\
                       &=& {1\over4i}([\gamma^M , \gamma^N]P_R)
                                    _{\alpha\beta}, \\
  \theta^{MN} &=& \Lm
    \begin{array}{cl}
      x_n & M+1=N=2n,~(n=1,\cdots,5) \\
      0   & {\rm otherwise},
    \end{array}
  \right.
\eeqn
where the $\gamma^M$ are $SO(10)$ $\gamma$-matrices and $P_R$ is the
right-handed projector, which can be written 
\beqn
  \gamma^1 &=& \tau_1 \otimes {\bf 1} \otimes {\bf 1}
                      \otimes {\bf 1} \otimes {\bf 1},  \\
  \gamma^2 &=& \tau_3 \otimes {\bf 1} \otimes {\bf 1}
                      \otimes {\bf 1} \otimes {\bf 1},  \\
  \gamma^3 &=& \tau_2 \otimes \tau_1 \otimes {\bf 1}
                      \otimes {\bf 1} \otimes {\bf 1},  \\
  \gamma^4 &=& \tau_2 \otimes \tau_3 \otimes {\bf 1}
                      \otimes {\bf 1} \otimes {\bf 1},  \\
  \gamma^5 &=& \tau_2 \otimes \tau_2 \otimes \tau_1
                      \otimes {\bf 1} \otimes {\bf 1},  \\
  \gamma^6 &=& \tau_2 \otimes \tau_2 \otimes \tau_3
                      \otimes {\bf 1} \otimes {\bf 1},  \\
  \gamma^7 &=& \tau_2 \otimes \tau_2 \otimes \tau_2
                      \otimes \tau_1 \otimes {\bf 1},  \\
  \gamma^8 &=& \tau_2 \otimes \tau_2 \otimes \tau_2
                      \otimes \tau_3 \otimes {\bf 1},  \\
  \gamma^9 &=& \tau_2 \otimes \tau_2 \otimes \tau_2
                      \otimes \tau_2 \otimes \tau_1,   \\
  \gamma^{10} &=& \tau_2 \otimes \tau_2 \otimes \tau_2
                         \otimes \tau_2 \otimes \tau_3,\\
  \gamma^{11} &=& i\gamma^1\gamma^2\gamma^3\gamma^4\gamma^5
                  \gamma^6\gamma^7\gamma^8\gamma^9\gamma^{10}\nn\\
              &=& \tau_2 \otimes \tau_2 \otimes \tau_2
                      \otimes \tau_2 \otimes \tau_2,   \\
  P_R &=& {1+\gamma^{11}\over2}.
\eeqn
In this basis, we have 
\beqn
  \theta^{MN}{T_{16}}^{MN} &=& -{1\over2}(
                    x_1 \tau_2 \otimes {\bf 1} \otimes {\bf 1}
                      \otimes {\bf 1} \otimes {\bf 1}  \nn\\
                &&\hspace{2mm}+ x_2 {\bf 1} \otimes \tau_2 \otimes {\bf 1}
                      \otimes {\bf 1} \otimes {\bf 1}  \nn\\
                &&\hspace{2mm}+ x_3 {\bf 1} \otimes {\bf 1} \otimes \tau_2
                      \otimes {\bf 1} \otimes {\bf 1}  \nn\\
                &&\hspace{2mm}+ x_4 {\bf 1} \otimes {\bf 1} \otimes {\bf 1}
                      \otimes \tau_2  \otimes {\bf 1}  \nn\\
                &&\hspace{2mm}+ x_5 {\bf 1} \otimes {\bf 1} \otimes {\bf 1}
                      \otimes {\bf 1} \otimes \tau_2
                                        ) P_R, \nn\\
                         &\equiv&  B, \\
  \theta^{MN}{T_{10}}^{MN} &=& \tau_2 \otimes {\rm diag}
                               (x_1,x_2,x_3,x_4,x_5).  \nn\\
                         &\equiv&  C.
\eeqn

  Before beginning the calculation, we should determine what coupling
can occur in the term $A^\prime A^3$ of $E_6$.
  Because ${\bf 78\times78 = 1_s+78_a+650_s+2430_s+2925_a}$,
$A^\prime A^3 \ni (A^\prime A)_{\bf 1}(A^2)_{\bf 1},
                  (A^\prime A)_{\bf 650}(A^2)_{\bf 650},
                  (A^\prime A)_{\bf 2430}(A^2)_{\bf 2430}$.
  On the other hand, because of the completeness,
\bequ
  (A_1A_2)_{\bf 2430}(A_3A_4)_{\bf 2430} =
               \sum_{I={\bf 1,78,650,2430,2925}}
               \lambda_I(A_1A_4)_I(A_3A_2)_I.
\eequ
  Therefore, 
\bequ
  (A^\prime A)_{\bf 2430}(A^2)_{\bf 2430} =
               \sum_{I={\bf 1,650,2430}}
               \lambda_I(A^\prime A)_I(A^2)_I,
\eequ
which implies that the above three couplings are not independent, 
and it is sufficient to examine the first two. They are essentially 
described as ${\rm tr}A^\prime A{\rm tr}A^2$ and 
${\rm tr}A^\prime A^3$ in matrix language.
If the desirable coupling existed, it would apparently be included
only in $(A^\prime A)_{\bf 650}(A^2)_{\bf 650}$ and
${\rm tr}A^\prime A^3$.
Thus we can conclude that it does not exist if
${\rm tr}A^\prime A^3$ does not include $\sum_i x^\prime_i {x_i}^3$.

  From (\ref{A-matrix}), we find 
\beqn
  {\rm tr}A^\prime A &=& {4\over3}y^\prime y
                        +{\rm tr}_{16}\Ll B^\prime B
                              + {1\over2\sqrt3}B^\prime y
                              + {1\over2\sqrt3}y^\prime B
                              + {1\over12}y^\prime y \Rl  \nn\\
                     &&   +{\rm tr}_{10}\Ll C^\prime C
                              - {1\over\sqrt3}C^\prime y
                              - {1\over\sqrt3}y^\prime C
                              + {1\over3}y^\prime y \Rl   \nn\\
                     &=& \Ls {4\over3}+{16\over12}+{10\over3}\Rs
                                                      y^\prime y
                        +\Ls 16{1\over2^2}+ 2\Rs
                                         \sum_i x^\prime_i x_i \nn\\
                     &=& 6\Ls y^\prime y + \sum_i x^\prime_i x_i\Rs.
\eeqn
  Similarly,
\beqn
  {\rm tr}A^\prime A^3 &=& {16\over9}y^\prime y^3
                          +{\rm tr}_{16}\Ll B^\prime B^3
                                + 3{1\over12}\Ls B^\prime By^2
                                + y^\prime yB^2\Rs
                                + {1\over144}y^\prime y^3 \Rl  \nn\\
                      &&    +{\rm tr}_{10}\Ll C^\prime C^3
                                + 3{1\over3}\Ls C^\prime Cy^2
                                + y^\prime yC^2\Rs
                                + {1\over9}y^\prime y^3 \Rl  \nn\\
                       &=& {16\over9}y^\prime y^3  
                       +16\Ll{1\over2^4}\Ls3\sum_i x^\prime_i x_i
                                             \sum_i x_i^2
                                            - 2\sum_i x^\prime_i
                                               x_i^3\Rs\right.\nn\\
                        &&       + \left.{3\over12}{1\over2^2}
                                           \Ls\sum_i x^\prime_ix_iy^2
                                            + y^\prime y \sum_i
                                              x^\prime_ix_i\Rs
                               + {1\over144}y^\prime y^3 \Rl  \nn\\
                      &&    +\Ll 2\sum_i x^\prime_ix_i^3
                                + {3\over3}\Ls 2\sum_i
                                                x^\prime_ix_iy^2
                                            + y^\prime y 2\sum_i
                                              x^\prime_ix_i\Rs
                                + {10\over9}y^\prime y^3 \Rl  \nn\\
                      &=& 3\Ls y^\prime y + \sum_i
                                  x^\prime_ix_iy^2  \Rs
                           \Ls y^2 + \sum_i x_i^2  \Rs  \nn\\
                      &=& {1\over12}{\rm tr}A^\prime A{\rm tr}A^2.
\eeqn
It is thus seen that desirable coupling does not exist because of 
the group theoretical cancellation between the contributions from 
the ${\rm tr}_{\rm 16}$ part and the ${\rm tr}_{\rm 10}$ part.\\

There are several solutions, and the simplest one is to use the term
${\overline \Phi}A^\prime A^3 \Phi$.
At first glance, it seems to have no effect, because
$\Phi {\overline \Phi}$ is written as
\begin{equation}
  \Phi {\overline \Phi} = \left(
  \begin{array}{ccc}
    \VEV{{\overline \Phi}\Phi} & 0 & 0 \\
    0 & 0_{16} & 0 \\
    0 & 0 & 0_{10} \\
  \end{array}
\right).
\end{equation}
  However this form is a special combination of
$(\Phi {\overline \Phi})_{\bf 1}, (\Phi {\overline \Phi})_{\bf 78}
$ and $(\Phi {\overline \Phi})_{\bf 650}$.
  In fact, we have 
\begin{eqnarray}
  \left(
  \begin{array}{ccc}
    \VEV{{\overline \Phi}\Phi} & 0 & 0 \\
    0 & 0_{16} & 0 \\
    0 & 0 & 0_{10} \\
  \end{array}
  \right) &=& {\VEV{{\overline \Phi}\Phi} \over 54} \Ll
  2\Ls
  \begin{array}{ccc}
    1 & 0 & 0 \\
    0 & {\bf1}_{16} & 0 \\
    0 & 0 & {\bf1}_{10} \\
  \end{array}
  \Rs + 3\Ls
  \begin{array}{ccc}
    4 & 0 & 0 \\
    0 & {\bf1}_{16} & 0 \\
    0 & 0 & -2\times{\bf1}_{10} \\
  \end{array}
  \Rs \right. \nn \\
  &&+ \left.\Ls
  \begin{array}{ccc}
    40 & 0 & 0 \\
    0 & -5\times{\bf1}_{16} & 0 \\
    0 & 0 & 4\times{\bf1}_{10} \\
  \end{array}
  \Rs\right],
\end{eqnarray}
where the three matrices on the rhs are proportional to the $SO(10)$ 
singlets of ${\bf 1}, {\bf 78}$ and ${\bf 650}$, respectively. 
Since the interactions for each 
representation have independent couplings, generically the cancellation
does not happen without fine-tuning.

There are several other solutions for this problem.
The essential ingredient is the interaction between $A'A^3$ and 
some other operator, whose VEV breaks $E_6$, 
because the cancellation
is due to a feature of the $E_6$ group. 
We now present some of these solutions.
\begin{itemize}
  \item  Allowing the higher-dimensional term $A^\prime A^5$. 
  Since $\VEV{A^2}$ breaks $E_6$, the cancellation can be avoided,
  which can be shown by a straightforward calculation.
  Since the number of solutions of the $F$-flatness conditions 
  increases, it becomes less natural to obtain a DW VEV. 
  But the number of vacua is still finite.
  \item  Introducing additional adjoint Higgs fields $B^\prime$ and 
       $B$, and giving $B$ the VEV pointing to a $SO(10)$-singlet.
         Then $B$ plays the same role as the above 
         $(\Phi {\overline \Phi})_{\bf 78}$.
         Examining the superpotential
   \begin{equation}
      W=B'B+\bar\Phi B'\Phi,
   \end{equation}
         the desired VEV $\VEV{{\bf 1}_B}\neq 0$ and 
         $\VEV{{\bf 45}_B}= 0$ is easily obtained.
\end{itemize}

\section{The VEV of an adjoint Higgs}  \label{D-flatness}
  In this appendix, we show that there is a gauge in which the 
VEV of an adjoint Higgs points in the direction of the 
Cartan subalgebra (CSA).

  Suppose that one Higgs $A$, which belongs to the adjoint 
representation of the group G (of dimension $d$ and rank  $r$), 
obtains a non-vanishing VEV.
  Then the D-flatness condition is as follows:
\beqn
  0   &=& \Ls A^b\Rs^\ast\Ls T_G^a\Rs_{bc}A^c  \nn\\
      &=& -i \Ls A^b\Rs^\ast f^{abc}A^c  \nn\\
  \Longrightarrow \hspace{1cm}
  0   &=& \Ls A^b\Rs^\ast f^{abc}A^c \Ls T^a\Rs\nn\\
      &=& \Ls A^b\Rs^\ast \Ll T^b , T^c\Rl A^c \nn\\
      &=& \Ll A^\dagger, A \Rl.                       \label{D-condi}
\eeqn
Next, we expand the VEV in the basis 
  $\{H^a, E_\alpha, E_{-\alpha}\}$.
  Here the $H^a$ $(a=1,\cdots,r)$ are contained in the
CSA, the values $\alpha$
are ${d-r\over2}$ positive roots, and $E_\alpha^\dagger
= E_{-\alpha}$.
  The following commutation relations hold: 
\beqn
  \Ll H^a , H^b \Rl   &=&   0,  \\
  \Ll H^a , E_{\pm\alpha} \Rl   &=&   \pm\alpha^aE_{\pm\alpha},  \\
  \Ll E_\alpha , E_{-\alpha} \Rl   &=&  \alpha^aH^a,  \\
  \Ll E_\alpha , E_{\pm\beta} \Rl   &=&  N_{\alpha,\pm\beta}
                                         E_{\alpha\pm\beta},
\eeqn
where the values $N_{\alpha,\pm\beta}$ are constants that depend on
$\alpha$ and $\beta$, which are nonzero only if $\alpha\pm\beta$ is 
also a root.
  In this basis, the VEV is written as
\begin{eqnarray}
  A   &=& A^a H^a + \sum_{\rm positive\ root\ \alpha}
                       \Lm A^\alpha_+ E_\alpha
                     + A^\alpha_- E_{-\alpha}\Rm,   \\
  A^\dagger &=& \Ls A^a\Rs^* H^a + \sum_\alpha
                       \Lm\Ls A^\alpha_-\Rs^* E_\alpha
                     + \Ls A^\alpha_+\Rs^*E_{-\alpha}\Rm .
\eeqn
  Then, extracting the part proportional to the CSA from the lhs of
(\ref{D-condi}), it becomes
\beqn
  \Ll A^\dagger, A\Rl &=& \sum_\alpha\Ls\left|A^\alpha_-\right|^2 -
                                    \left|A^\alpha_+\right|^2 \Rs
                                 {\bf \alpha}^aH_a
                           + \cdots  \nn\\
                       &=&  0.
\end{eqnarray}
  Therefore, if all the values $A^\alpha_-$ are zero, then $\sum_\alpha
\left|A^\alpha_+\right|^2{\bf \alpha}^a$ is zero, and therefore all
the values $A^\alpha_-$ are also zero.%
\footnote{  Though even positive roots may have negative components,
          such roots must have a positive component at smaller
          values of $a$ by the definition of a positive root. 
            Therefore, examining the conditions from that of a smaller 
          value of $a$ to that of larger one successively, 
          $\sum_\alpha\left|A^\alpha_+\right|^2{\bf \alpha}^a=0$ is 
          easily confirmed.
}

  Now, we show that all the $A^\alpha_-$ can be rotated away
through the gauge rotation.
  For infinitesimal rotations, the transformation law of $A$ is
\bequ
  -i\delta A  = \Ll \theta^a H^a + \sum_\alpha
                         \Lm \theta^\alpha_+ E_\alpha
                       + \theta^\alpha_- E_{-\alpha}\Rm,
                     A^a H^a + \sum_\alpha
                         \Lm A^\alpha_+ E_\alpha
                       + A^\alpha_- E_{-\alpha}\Rm \Rl ,  \nn
\eequ
where $\theta$ is the parameter of the rotation composed of
$d$ real numbers corresponding to $d$ generators.
  Then, extracting the part proportional to $E_\alpha$, we have 
\bequ
  -i\delta A^\alpha_- = \Ls A^a \theta^\alpha_-
                          - A^\alpha_- \theta^a \Rs \alpha^a
                      +\sum_{\beta-\gamma=-\alpha}
                          \Ls A^\gamma_- \theta^\beta_+
                            - A^\beta_+ \theta^\gamma_- \Rs
                          N_{\beta,-\gamma}
                       -\sum_{\beta+\gamma=\alpha}
                          A^\gamma_- \theta^\beta_-
                          {N_{\beta,\gamma}}^\ast .
\eequ
  It is seen that there are $d-1$ gauge degrees of freedom, 
except for the one corresponding to the maximum root, 
compared with the ${d-r\over2}$ complex VEVs corresponding to 
$d-r$ real numbers.
  Therefore, all the $A^\alpha_-$ can be rotated away, and
in this gauge, the D-flatness condition forces all the $A^\alpha_+$
to be zero.

  Now, we have shown that the VEV of an adjoint Higgs can be expressed
as pointing to the CSA.
  In other words, the VEV of an adjoint Higgs is gauge equivalent to
that pointing to the CSA in the supersymmetric limit.

\section{Operators that induce mass matrices}
In this appendix, we give the operators that induce the mass matrices
of superheavy particles in $E_6$ GUT.

First, we examine the operator matrix $O_{24}$ 
of ${\bf 24}$ in $SU(5)$, which induces the mass matrices
$M_I~(I=X,G,W)$, 
\begin{equation}
O_{24}=\bordermatrix{
\bar I\backslash I  &    24_A(-1)   &     24_{A'}(4)           \cr
24_A(-1)                &     0      &      A'A \cr
24_{A'}(4)   &  A'A   &  {A'}^2       \cr
},
\end{equation}
where the numbers in the parentheses denote typical charges.

Next, we examine the operator matrix $O_{10}$ of ${\bf 10}$ in $SU(5)$,
which induces the mass matrices $M_I~(I=Q, U^c, E^c)$, 
\begin{equation}
\bordermatrix{
I\backslash \bar I  &\overline {16}_{\bar \Phi}(2) & 
\overline {16}_{\bar C}(-2)&\overline {16}_{A}(-1) &45_{A}(-1) &
\overline {16}_{\bar C'}(8)   & 
                    \overline {16}_{\bar A'}(4) &  45_{A'}(4)  \cr
16_\Phi(-3) & 0 & 0 & 0& 0& \bar C'A\Phi  & \bar\Phi A'A\Phi
         & 0 \cr
16_C(-6)& 0 & 0 &0 & 0& \bar C'AC &  0  & 0 \cr
16_A(-1) & 0 & 0 & 0& 0& \bar C'A\Phi   & \bar \Phi A'A\Phi
   & 0 \cr
45_{A}(-1) & 0 & 0 & 0 & 0& \bar C'AC  & 0 & 
       A'A  \cr
16_{C'}(7) & \bar\Phi A C' & \bar CAC' &
        \bar \Phi AC' & \bar CAC' & 
        \bar C'C' & \bar\Phi A'C' & 
        \bar CA'C'  \cr
16_{A'}(4) & \bar \Phi A'A\Phi & 0 & 
        \bar\Phi A'A\Phi & 0&\bar C'A'\Phi &  
        {A'}^2& \bar C{A'}^2\Phi  \cr
45_{A'}(4) & 0 & 0 &0 & A'A
        & \bar C'A'C &  
        \bar\Phi {A'}^2C 
        & {A'}^2 \cr
},
\end{equation}
where we have given only one example, even if there are several 
corresponding operators.

Finally, we examine the operator matrix $O_5$ of ${\bf 5}$ and ${\bf \bar 5}$
in $SU(5)$, which induces the mass matrices $M_I~(I=L,D^c)$, 
\begin{equation}
O_5=\left(\matrix{ 0 & A_5 & 0 \cr  B_5 & C_5 & D_5 \cr 
                                    E_5 & F_5 & G_5 \cr}\right),
\end{equation}
\begin{equation}
A_5=\bordermatrix{
\bar I\backslash I   &  10_{C'}(7)
                    & 10_{\bar C'}(8) &
                    \overline{16}_{\bar C'}(8) &
                    \overline{16}_{A'}(4)  \cr
10_\Phi(-3) & C'A\Phi^2 & \bar C'(A+Z)\Phi & 
0  & 0 \cr
10_C (-6) & 0 & \bar C'(A+Z)C & 0 & 0 \cr
16_C(-6) & 0 & 0 & \bar C'(A+Z)C
   & 0 \cr
16_A(-1)  & 0 & \bar C'AC & \bar C'A\Phi
   & A'A \cr
}, 
\end{equation}
\begin{equation}
B_5=\bordermatrix{
\bar I\backslash I  & 10_\Phi(-3)&10_C (-6)&\overline{16}_{\bar C}(-2)&
\overline{16}_{A}(-1)\cr
10_{C'}(7) & C'A\Phi^2 & 0 & 0 & \bar CAC' \cr
10_{\bar C'}(8) &\bar C'(A+Z)\Phi &\bar C'(A+Z)C &
\bar C'(A+Z)\bar C^2 & \bar C'A\bar\Phi\bar C \cr
16_{C'}(7) & 0 & 0 &\bar C(A+Z)C' & \bar\Phi AC' \cr
16_{A'}(4)& 0 & 0 & 0 & A'A \cr
}, 
\end{equation}
\begin{equation}
C_5=\bordermatrix{
\bar I\backslash I  & 10_{C'} (7)
                    & 10_{\bar C'}(8)  &
                    \overline{16}_{\bar C'}(8) &
                    \overline{16}_{A'}(4)  \cr
10_{C'}(7)  & {C'}^2\Phi & 
  \bar C'C' &
  \bar C'\bar C C'\Phi  & \bar CA'C' \cr
10_{\bar C'}(8)  &\bar C'C' & 
(\bar C')^2\bar \Phi & (\bar C')^2\bar C & 
\bar C'A'\bar C\bar\Phi \cr
16_{C'} (7)& {C'}^2C & 
  \bar C'\bar\Phi C'C & \bar C'C' &  
\bar\Phi A'C' \cr
16_{A'}(4)& C'A'\Phi C & 
\bar C'A'C & \bar C'A'\Phi &  
{A'}^2 \cr
},
\end{equation}
\begin{equation}
D_5=\bordermatrix{
\bar I\backslash I  & 10_{\bar \Phi}(2) &10_{\bar C}(-2) &
\overline{16}_{\bar \Phi}(2) \cr
10_{C'}(7) &\bar\Phi(A+Z)C' & \bar C(A+Z)C' & 
   \bar \Phi\bar C C'\Phi \cr
10_{\bar C'}(8) & \bar C'A\bar\Phi^2 & 
       \bar C'A\bar C\bar \Phi & 
       \bar C'A\bar\Phi\bar C \cr
16_{C'} (7)& \bar\Phi^2AC'C &
      \bar C\bar\Phi AC'C & \bar\Phi(A+Z)C' \cr
16_{A'} (4)& 0 & 0 & \bar\Phi A'A\Phi \cr
}, 
\end{equation}
\begin{equation}
E_5=
\bordermatrix{
\bar I\backslash I  &10_\Phi(-3) & 10_C(-6)& 
                    \overline{16}_{\bar C}(-2) &\overline{16}_{A}(-1)   \cr
10_{\bar \Phi}(2) & 
0 & 0& 0 & 
\bar\Phi^2A^2\bar C   \cr
10_{\bar C} (-2)&
0 & 0& 0 & 0
 \cr
16_\Phi (-3)& 0 & 0 &  0 & 0  \cr
},
\end{equation}
\begin{equation}
F_5=\bordermatrix{
\bar I\backslash I   &  10_{C'} (7)
                    & 10_{\bar C'} (8)&
                    \overline{16}_{\bar C'} (8)&
                    \overline{16}_{A'}(4)  \cr
10_{\bar \Phi} (2) &
\bar\Phi(A+Z)C' & \bar C'A\bar\Phi^2 &
\bar C'A\bar\Phi\bar C & 
 \bar \Phi^2A'\bar C \cr
10_{\bar C} (-2) & \bar C(A+Z)C' &
\bar C'A\bar C\bar\Phi& \bar C'A\bar C^2 & 
\bar C^2A'A\bar\phi \cr
16_\Phi (-3) & 0 & \bar C'A\bar\Phi C\Phi& 
   \bar C'(A+Z)\Phi
   & \bar\Phi A'A\Phi \cr
}, 
\end{equation}
\begin{equation}
G_5=\bordermatrix{
\bar I\backslash I   &  10_{\bar \Phi} (2)
                    & 10_{\bar C}(-2) &
                    \overline{16}_{\bar \Phi} (2)\cr
10_{\bar \Phi} (2) &
\bar\Phi^3 & \bar\Phi^2\bar C &
\bar\Phi^2\bar C \cr
10_{\bar C} (-2) & \bar\Phi^2\bar C & 0 & 0 \cr
16_\Phi (-3) & 0 & 0 & 0 \cr
}.
\end{equation}

\end{document}